\documentclass[12pt]{article}
\usepackage{qSym}
\input{MyQcircuit.sty}
\begin{document}
\title{Quantum Circuit for Calculating\\
Symmetrized Functions\\
Via Grover-like Algorithm}

\author{Robert R. Tucci\\
        P.O. Box 226\\
        Bedford,  MA   01730\\
        tucci@ar-tiste.com}

\date{\today}
\maketitle
\vskip2cm
\section*{Abstract}
In this
paper, we give
a quantum circuit that
calculates
symmetrized functions.
Our algorithm
applies
the original Grover's algorithm
or a variant thereof such as
AFGA (adaptive fixed point Grover's algorithm).
Our algorithm uses AFGA in conjunction with
two new techniques we call ``targeting
two hypotheses" and ``blind targeting".
Suppose AFGA drives the starting state $|s\rangle$
to the target state $|t\rangle$.
When targeting two hypotheses,
$|t\rangle$ is a superposition $a_0|0\rangle + a_1|1\rangle$ of
two orthonormal states or hypotheses $\ket{0}$ and $\ket{1}$.
When targeting blindly,
the value of $\langle t| s\rangle$
is not known a priori.

\newpage

\section{Introduction}

In this
paper, we give
a quantum circuit that
calculates
symmetrized functions
(i.e.,
it calculates
the right hand
side of Eq.(\ref{eq-goal})).

Our algorithm
utilizes
the original Grover's algorithm (see
Ref.\cite{Gro})
or any variant thereof,
as long as it
accomplishes the task
of driving
a starting state
$\ket{s}$  towards a target state
$\ket{t}$.
However, we recommend
to the users of our algorithm
that
they use
a variant
of Grover's algorithm
called AFGA (adaptive
fixed point Grover's algorithm)
which was first
proposed in Ref.\cite{afga}.

A large portion of
our algorithm for
calculating symmetrized functions
has been
proposed before by Barenco et al
in Ref.\cite{Bar}.
However,
we make some important changes to
their algorithm.
One trivial
difference between our work
and that of Barenco et al
is that our operators
$V^{(\lam)}_1$
are different from
the corresponding ones that
Barenco et al use.
A more important difference
is that we combine
their circuit with
Grover's algorithm (or
variant thereof),
which they don't.
Furthermore,
we use Grover's algorithm
in conjunction
with two new techniques that we call
``targeting two hypotheses"
and ``blind targeting".
When targeting two hypotheses,
$|t\rangle$ is a superposition $a_0|0\rangle + a_1|1\rangle$ of
two orthonormal states or hypotheses
$\ket{0}$ and $\ket{1}$.
When targeting blindly,
the value of $\langle t| s\rangle$
is not known a priori.

The technique of ``targeting two hypotheses"
can be used in conjunction with Grover's algorithm or variants thereof to
estimate (i.e., infer) the
amplitude of one of many states in a superposition.
An earlier technique
by Brassard et al (Refs.\cite{Bra1,Bra2})
can also be used in conjunction with Grover's algorithm to achieve the same goal of amplitude inference.
However, our technique is very different
from that of Brassard et al.
They try to
produce a ket $\ket{x^n}$,
where the bit string $x^n$
encodes the amplitude
that they are trying to infer.
We, on the other hand,
try to infer an amplitude $|a_1|$
by measuring the ratio $|a_1|/|a_0|$
and assuming we know $|a_0|$ a priori.

For more background
information on the use
of symmetrized functions
in quantum information theory,
we refer the reader to
a recent review by Harrow, Ref.\cite{Har}.

\section{Notation and Preliminaries}

In this section,
we will review
briefly some
of the more
unconventional
notation used
in this paper.
For a more detailed
discussion of Tucci's
notation,
especially its
more idiosyncratic aspects,
see, for example,
Ref.\cite{Paulinesia}.

Let
$\theta({\cal S})$
stand for the truth function. It
equals
1 when statement ${\cal S}$ is true
and 0 when it isn't. The Kronecker
delta function $\theta(a=b)$
will also be denoted by
$\delta_a^b$ or $\delta(a,b)$.
Given a set $A$, the indicator
function for set $A$ is defined by
$1_A(x) = \theta(x\in A)$.

We will sometimes use the
following abbreviation
for sets:
$\{f(x):\forall x\in S\}=\{f(x)\}_{\forall x}$.

We will sometimes use the following
abbreviation for Hermitian conjugates:
$[x] + [h.c.] = x + x^\dagger$,
and
$[x][h.c.] = x x^\dagger$,
where $x$ is some complicated
expression that we don't want to write twice.

We will sometimes use the following
abbreviation:
$\frac{f(x)}{\sum_x num}=
\frac{f(x)}{\sum_x f(x)}$,
where $f(x)$ is some complicated
expression of $x$
that we don't want to write twice.

Let
$Bool=\{0,1\}$.
For any $b\in Bool$, let
$\olb=1-b$.

Let $\CC$ stand for the complex numbers and $\RR$ for the real numbers. For integers $a,b$ such that
$a\leq b$, let
$\{a..b\}=\{a,a+1,\ldots, b\}=\{b..a\}$.

We will
represent $n$-tuples or vectors with $n$ components by $x^n=x.=(x_{n-1}, x_{n-2}, \ldots , x_1, x_0)$. If $x^n=(x_j)_{\forall j}\in Bool^n$, then
define functions $dec()$ and $bin^n()$ by
$dec(x^n) = \sum_{j=0}^{n-1} 2^j x_j$ and
$bin^n(\sum_{j=0}^{n-1} 2^j x_j)= x^n$.

We will use the
 term qu(d)it to refer to a quantum
 system that lives in a $d$-dimensional Hilbert space $\CC^d =span_\CC\{\ket{j}: j=0,1,\ldots,d-1\}$.
Hence a qu(4)it has 4 possible
independent states. A qubit is a qu(2)it.
Systems (or
horizontal wires in a quantum
circuit) will
be labelled by Greek letters.
If $\alpha$
lives in the Hilbert space
$(\CC^d)^{\otimes n}$,
we will say $width(\alpha)=d^n$.
For example, we'll say
$width(\alpha) = 3^5$ if wire $\alpha$ carries 5 qu(3)its.

As is usual
in the Physics literature, $\sigma_X,\sigma_Y, \sigma_Z$
will denote the Pauli matrices.
$H=\frac{1}{\sqrt{2}}\left[ \begin{array}{cc}
1&1
\\
1&-1
\end{array}\right]$
will denote the 1 qubit
Hadamard matrix.
$H^{\otimes n}$, the $n$-fold tensor product
of $H$,
is the $n$ qubits Hadamard matrix.

Define the number
operator $n$ and its
complement $\overline{n}$ by

\beq
n=P_1 = \ket{1}\bra{1}
\;,\;\;
\ol{n} = 1-n = P_0= \ket{0}\bra{0}
\;.
\eeq
If we need to distinguish
the number operator
from an integer called $n$,
we will use $n_{op}$
or $\hat{n}$, or $\ul{n}$
for the number operator.

The number operator just defined
acts only on qubits. For qu(d)its,
one can use instead

\beq
P_b=
\ket{b}\bra{b}
\;,
\eeq
where
$b\in\{0,1,\ldots,d-1\}$.
For 2 qu(d)its, one can use

\beq
P_{b_1}\otimes P_{b_0}=P_{b_1}(\beta_1)
P_{b_0}(\beta_0)=
P_{b_1,b_0}(\beta_1,\beta_0)
\;,
\label{eq-pb-2}
\eeq
where
$b_1,b_0\in\{0,1,\ldots,d-1\}$.
Eq.(\ref{eq-pb-2})
generalizes easily to an arbitrary number of
qu(d)its.

We will often denote
tensor products of kets
vertically instead of horizontally.
The horizontal and vertical notations
will be related by the conventions:

\beq
\ket{a_{n-1}}\otimes\ldots
\ket{a_1}\otimes
\ket{a_0}
=
\ket{a_{n-1},\ldots,a_1,a_0}
=
\begin{array}{c}
\ket{a_0}
\\
\ket{a_1}
\\
\vdots
\\
\ket{a_{n-1}}
\end{array}
\;
\eeq
and

\beq
(\ket{a_{n-1},\ldots,a_1,a_0})^\dagger
=
\bra{a_{n-1},\ldots,a_1,a_0}
=
\begin{array}{c}
\bra{a_0}
\\
\bra{a_1}
\\
\vdots
\\
\bra{a_{n-1}}
\end{array}
\;.
\eeq

As usual for us, we will
represent various types of
controlled nots
as follows:

\beq
\begin{array}{c}
\Qcircuit @C=1em @R=1em @!R{
&\dotgate&\qw&\scriptstyle{\alpha}
\\
&\timesgate\qwx[-1]&\qw&\scriptstyle{\beta}
}
\end{array}
=
\sigma_X(\beta)^{n(\alpha)}
\;,\;\;
\begin{array}{c}
\Qcircuit @C=1em @R=1em @!R{
&\ogate&\qw&\scriptstyle{\alpha}
\\
&\timesgate\qwx[-1]&\qw&\scriptstyle{\beta}
}
\end{array}
=
\sigma_X(\beta)^{\ol{n}(\alpha)}
\;,\;\;
\begin{array}{c}
\Qcircuit @C=1em @R=1em @!R{
&\dotgate&\qw&\scriptstyle{\alpha_0}
\\
&\dotgate&\qw&\scriptstyle{\alpha_1}
\\
&\timesgate\qwx[-2]&\qw&\scriptstyle{\beta}
}
\end{array}
=
\sigma_X(\beta)^{n(\alpha_0)n(\alpha_1)}
\eeq

We will represent as follows a
controlled $U$,
where the unitary operator $U$
acts
on $\alpha$
and where $\beta$
is the control:

\beq
\begin{array}{c}
\Qcircuit @C=1em @R=1em @!R{
&\gate{U}&\qw&\scriptstyle{\alpha}
\\
&\dotgate\qwx[-1]&\qw&\scriptstyle{\beta}
}
\end{array}
\;\;\;\;
=
U(\alpha)^{n(\beta)}
\;.
\eeq
Note that
$
[U(\alpha)^{n(\beta)}][h.c.]=1
$
so controlled unitaries
are themselves unitary.

We will use the following
identity repeatedly throughout this paper. For any
quantum systems $\alpha$ and $\beta$,
any
unitary operator $U(\beta)$
and any
projection operator $\pi(\alpha)$
(i.e., $\pi^2=\pi$),
one has

\beq
U(\beta)^{\pi(\alpha)}=
(1-\pi(\alpha)) + U(\beta)\pi(\alpha)
\;.
\label{eq-u-pi-id}
\eeq

We will denote ordered
products of operators $U_b$
as follows:
\beq
\prod_{b=0\rarrow2} U_b=
U_0 U_1 U_2
\;,\;\;
\prod_{b=2\rarrow0} U_b=
U_2 U_1 U_0
\;.
\eeq

Suppose $a,b\in Bool$ and $x,y,\theta$ are real numbers. Note that
$\delta_a^0 = \ola$ and $\delta_a^1=a$.
Furthermore, note that
$x^a y^\ola= xa + y\ola= x(a)$ where $x(a) = x$ if $a=1$ and $x(a)=y$ if $a=0$.
If we let $S = \sin \theta$ and $C= \cos \theta$, then

\beqa
\av{a | e^{-i\sigma_Y \theta b} |0}
&=&
\delta_a^0 \delta _b^0 +
(C \delta_a^0 + S \delta_a^1)\delta_b^1
\\
&=& \ola \olb + (C \ola + S a)b
\;.
\eeqa

\section{Permutation Circuits}

For this section,
we will assume that the reader
has a rudimentary
knowledge of permutations,
as can be obtained
from any first
course in abstract algebra.
In this section,
we will attempt to
connect that rudimentary knowledge
of permutations
with quantum computation.
More specifically,
we will show
how to permute
the  qu(d)its
of a multi-qu(d)it
quantum state
using a quantum circuit.

Given any finite set $S$,
a permutation on set $S$
is a 1-1 onto map from $S$ to $S$.

Define

\beq
Sym(S)= \{ \sigma| \sigma\mbox{ is a permutation of set } S\}
\;.
\eeq
The properties of
$Sym(S)$ don't depend on the nature
of $S$, except for its
cardinality $|S|$ (i.e., number of elements
of $S$). Hence, we will
often denote $Sym(S)$ by
$Sym_{|S|}$.

If permutation $\sigma$ maps
$x\in S$ to $\sigma(x)\in S$,
we will often write $\sigma(x) = x^\sigma$.
For example, if $\sigma$ maps 1 to 2, we
will write $1^\sigma = 2$.

As usual, a permutation $\sigma$
will be represented by

\beq
\left(
\begin{array}{cccc}
1&2&\cdots&n-1
\\
1^\sigma &2^\sigma &\cdots & (n-1)^\sigma
\end{array}
\right)=
(1^\sigma, 2^\sigma,\dots, (n-1)^\sigma)
\;
\eeq

For $\sigma\in Sym(S)$,
and any set $A$, define

\beq
A^\sigma=
\{a^{\sigma'}: a\in A\}\mbox{, where } a^{\sigma'}
=\left\{
\begin{array}{l}
a^\sigma \mbox{ if } a\in S
\\
a \mbox{ if } a\notin S
\end{array}
\right.
\;.
\eeq
For example,
if $S=\{1,2,3\}$ then
$\{1,2,4\}^\sigma =
\{1^\sigma, 2^\sigma, 4\}$.

If $\sigma\in Sym_n$
and $c^{n} =(c_{(n-1)},
\ldots, c_{1},c_{0})\in (S_\rvc)^n$,
define

\beq
c^{n\sigma} =(c_{(n-1)}^\sigma,
\ldots, c_{1}^\sigma,c_{0}^\sigma)
=
(c_{(n-1)^\sigma},
\ldots, c_{1^\sigma},c_{0^\sigma})
\;.
\eeq

For any
permutation map $\sigma:S\rarrow S$,
one can define a matrix
such that each of its
columns has all entries equal to zero
except for one single entry which
equals
1.
Also, the entry that is
1 is at a different
position for each column.
We will denote such a
matrix (which is orthogonal
and unitary)
also by $\sigma$.
Whether
$\sigma$ stands for
the map or the matrix
will be clear from context,
as in the following equation
which uses $\sigma$ to stand
for the matrix on its left side
and the map on its right side:

\beq
\sigma
\begin{array}{l}
\ket{a_0}
\\
\ket{a_1}
\\
\vdots
\\
\ket{a_{n-1}}
\end{array}
=
\begin{array}{l}
\ket{a_{0^\sigma}}
\\
\ket{a_{1^\sigma}}
\\
\vdots
\\
\ket{a_{(n-1)^\sigma}}
\end{array}
\;
\eeq

Suppose $a^n=(a^{n-1}, a^{n-2},\ldots, a^0)\in (S_\rva)^n$
and $\av{b^n|a^n}=\delta_{a^n}^{b^n}$
for all $a^n,b^n\in S^n_\rva$.
If $|S_\rva|=d$, then
we can assume without loss of generality that  $S_\rva = \{0..d-1\}$.
Suppose
$width(\alpha^n)=d^n$. Let

\beq
\ket{\psi}_{\alpha^n}=
\sum_{a^n}A(a^n)\ket{a^n}_{\alpha^n}
\;,\;\;
A(a^n) =\av{a^n|\psi}
\;.
\eeq
Then

\beq
\begin{array}{c}
\bra{a_0}
\\
\bra{a_1}
\\
\vdots
\\
\bra{a_{n-1}}
\end{array}
\sigma
\ket{\psi}_{\alpha^n}
=
\begin{array}{c}
\bra{a_{0^\tau}}
\\
\bra{a_{1^\tau}}
\\
\vdots
\\
\bra{a_{(n-1)^\tau}}
\end{array}
\ket{\psi}
=
A(a^{n\tau})
\;,
\eeq
where $\tau=\sigma^{-1}$.
When $\sigma$
is a permutation matrix, it's unitary so
$\sigma^{-1}=\sigma^\dagger$.

Define
\beq
\pi_{Sym_n}=\frac{1}{n!}\sum_{\sigma\in Sym_n}\sigma
\;.
\eeq
One finds that

\beq
[\pi_{Sym_n}]^2 = \pi_{Sym_n}
\;
\eeq
so $\pi_{Sym_n}$ is a projection
operator.
Furthermore,
one finds that

\beq
\av{a^n|\pi_{Sym_n}|\psi}=
\frac{1}{n!}
\sum_\sigma A(a^{n\sigma})
\;.
\label{eq-goal}
\eeq
The goal of this paper
is to find a quantum circuit
that allows us to
calculate $|\av{a^n|\pi_{Sym_n}|\psi}|^2$
for some predetermined point $a^n\in \{0..d-1\}^n$ and
state $\ket{\psi}_{\alpha^n}$,
where $width(\alpha^n)=d^n$.

As is well known,
any permutation can
be expressed as a product
of transpositions (a.k.a. swaps).
For quantum circuits,
it is common to define
a swap gate which acts as follows:

 \beq
 \begin{array}{c}
\Qcircuit @C=1em @R=1em @!R{
&\lamgate&\qw\;\;\;\;\;\ket{\psi_1}
\\
&\veegate\qwx[-1]&\qw\;\;\;\;\;\ket{\psi_2}
\\
&\qw&\qw\;\;\;\;\;\ket{\psi_3}
}
\end{array}
\;\;\;\;=
\begin{array}{c}
\Qcircuit @C=1em @R=1em @!R{
&\qw&\qw\;\;\;\;\;\ket{\psi_2}
\\
&\qw&\qw\;\;\;\;\;\ket{\psi_1}
\\
&\qw&\qw\;\;\;\;\;\ket{\psi_3}
}
\end{array}
\;.
\eeq
In this example, the gate $swap(1,2)$ is acting on 3 qu(d)its called 1,2,3.
Clearly, $
\left[
\begin{array}{c}
\Qcircuit @C=1em @R=1em @!R{
&\lamgate&\qw
\\
&\veegate\qwx[-1]&\qw
}
\end{array}
\right]^2=1
$.
One also finds that

\beqa
\left[
\begin{array}{c}
\Qcircuit @C=1em @R=1em @!R{
\wedge&\lamgate&\lamgate
\\
\vee\qwx[-1]&\qw&\veegate\qwx[-1]
\\
&\veegate\qwx[-2]&\qw
}
\end{array}
=
\begin{array}{c}
\Qcircuit @C=1em @R=1em @!R{
&\qw&\qw
\\
&\lamgate&\qw
\\
&\veegate\qwx[-1]&\qw
}
\end{array}
\right]
\;,\;\;
\left[
\begin{array}{c}
\Qcircuit @C=1em @R=1em @!R{
&\lamgate&\qw
\\
\wedge&\veegate\qwx[-1]&\lamgate
\\
\vee\qwx[-1]&\qw&\veegate\qwx[-1]
}
\end{array}
=
\begin{array}{c}
\Qcircuit @C=1em @R=1em @!R{
&\lamgate&\qw
\\
&\qw&\qw
\\
&\veegate\qwx[-2]&\qw
}
\end{array}
\right]
\;,\;\;
\left[
\begin{array}{c}
\Qcircuit @C=1em @R=1em @!R{
&\lamgate&\qw
\\
\wedge&\qw&\lamgate
\\
\vee\qwx[-1]&\veegate\qwx[-2]&\veegate\qwx[-1]
}
\end{array}
=
\begin{array}{c}
\Qcircuit @C=1em @R=1em @!R{
&\lamgate&\qw
\\
&\veegate\qwx[-1]&\qw
\\
&\qw&\qw
}
\end{array}
\right]
\;.
\eeqa
One can summarize
these 3 identities by saying that
the horizontal line with
{\it 3 arrow heads} on it
can be replaced by {\it no arrow heads} on it.
At the same time, the horizontal
line with
{\it 2 arrow heads} on it
can be replaced by {\it 1 arrow head} on it.

Note that the elements of $Sym_3$ in the so called dictionary order are

\beq
\begin{array}{cccccc}
\begin{array}{c}
(1)
\\
(1,2,3)
\\
\Qcircuit @C=1em @R=1em @!R{
\scriptscriptstyle{1}\;\;&\qw
\\
\scriptscriptstyle{2}\;\;&\qw
\\
\scriptscriptstyle{3}\;\;&\qw
}
\end{array}
&
\begin{array}{c}
(2)
\\
(1,3,2)
\\
\Qcircuit @C=1em @R=1em @!R{
&\qw
\\
&\lamgate
\\
&\veegate\qwx[-1]
}
\end{array}
&
\begin{array}{c}
(3)
\\
(2,1,3)
\\
\Qcircuit @C=1em @R=1em @!R{
&\lamgate
\\
&\veegate\qwx[-1]
\\
&\qw
}
\end{array}
&
\begin{array}{c}
(4)
\\
(2,3,1)
\\
\Qcircuit @C=1em @R=1em @!R{
\wedge\qwx[2]
&\lamgate
\\
&\veegate\qwx[-1]
\\
\vee& \qw
}
\end{array}
&
\begin{array}{c}
(5)
\\
(3,1,2)
\\
\Qcircuit @C=1em @R=1em @!R{
\wedge&\lamgate
\\
\vee\qwx[-1]&\qw
\\
&\veegate\qwx[-2]
}
\end{array}
&
\begin{array}{c}
(6)
\\
(3,2,1)
\\
\Qcircuit @C=1em @R=1em @!R{
&\lamgate
\\
&\qw
\\
&\veegate\qwx[-2]
}
\end{array}
\end{array}
\;.
\label{eq-sym3-dict}
\eeq

Note that the sum of the
6 elements of $Sym_3$
can be generated from
a product of matrices
which are themselves sums
of permutation matrices, as follows:

\beqa
\lefteqn{
\left(
\begin{array}{c}
\Qcircuit @C=1em @R=1em @!R{
&\qw
\\
&\qw
\\
&\qw
}
\end{array}
+
\begin{array}{c}
\Qcircuit @C=1em @R=1em @!R{
&\qw
\\
&\lamgate
\\
&\veegate\qwx[-1]
}
\end{array}
+
\begin{array}{c}
\Qcircuit @C=1em @R=1em @!R{
&\lamgate
\\
&\qw
\\
&\veegate\qwx[-2]
}
\end{array}
\right)
\left(
\begin{array}{c}
\Qcircuit @C=1em @R=1em @!R{
&\qw
\\
&\qw
\\
&\qw
}
\end{array}
+
\begin{array}{c}
\Qcircuit @C=1em @R=1em @!R{
&\lamgate
\\
&\veegate\qwx[-1]
\\
&\qw
}
\end{array}
\right)
=}
\nonumber
\\
&=&
\begin{array}{c}
(1)
\\
\Qcircuit @C=1em @R=1em @!R{
&\qw
\\
&\qw
\\
&\qw
}
\end{array}
+
\begin{array}{c}
(2)
\\
\Qcircuit @C=1em @R=1em @!R{
&\qw
\\
&\lamgate
\\
&\veegate\qwx[-1]
}
\end{array}
+
\begin{array}{c}
(6)
\\
\Qcircuit @C=1em @R=1em @!R{
&\lamgate
\\
&\qw
\\
&\veegate\qwx[-2]
}
\end{array}
+
\begin{array}{c}
(3)
\\
\Qcircuit @C=1em @R=1em @!R{
&\lamgate
\\
&\veegate\qwx[-1]
\\
&\qw
}
\end{array}
+
\begin{array}{c}
(5)
\\
\Qcircuit @C=1em @R=1em @!R{
\wedge&\lamgate
\\
\vee\qwx[-1]&\qw
\\
&\veegate\qwx[-2]
}
\end{array}
+
\begin{array}{c}
(4)
\\
\Qcircuit @C=1em @R=1em @!R{
\wedge\qwx[2]
&\lamgate
\\
&\veegate\qwx[-1]
\\
\vee& \qw
}
\end{array}
\nonumber
\\
&=&\sum_{\sigma\in Sym_3}\sigma
\;.
\label{eq-sym3-prod-sum}
\eeqa

It's fairly clear how to generalize
the pattern of
Eq. (\ref{eq-sym3-prod-sum})
to the case of $n$ qu(d)its and
$Sym_n$, where $n$ is any integer greater
than 1.

\section{Decomposing a State Vector into 2 Orthogonal Projections}

In this section, we will
review a technique that we like to
call
``decomposing a state
vector into orthogonal projections".
This technique is
frequently used in quantum
computation circuits, and
will be used later on in this paper,
inside more complicated circuits.

Suppose
$\alpha$
is a qu(d)it and $\beta$
is a qubit. Let
$\pi$ be a Hermitian projection operator
(i.e., $\pi^\dagger = \pi$, $\pi^2=\pi$)
acting on $\alpha$,
and let $\ol{\pi}=1-\pi$.
Let $\ket{\psi}_\alpha$
be a state vector of qu(d)it $\alpha$.
Applying identity Eq.(\ref{eq-u-pi-id}) with $U=\sigma_X(\beta)$
yields:

\beq
\sigma_X(\beta)^{\pi(\alpha)}
\begin{array}{l}
\ket{\psi}_\alpha
\\
\ket{0}_\beta
\end{array}
=
\sigma_X(\beta)^{\ol{\pi}(\alpha)}
\begin{array}{l}
\ket{\psi}_\alpha
\\
\ket{1}_\beta
\end{array}
=
\begin{array}{c}
\ol{\pi}(\alpha)\ket{\psi}_\alpha
\\
\ket{0}_\beta
\end{array}
+
\begin{array}{c}
\pi(\alpha)\ket{\psi}_\alpha
\\
\ket{1}_\beta
\end{array}
\;.
\eeq
One can say that the
state vector
$\ket{\psi}$ is ``decomposed"
by the circuit
into two orthogonal projections
$\pi\ket{\psi}$ and
$\ol{\pi}\ket{\psi}$.
Some examples of this
decomposition are (1)
when $\alpha$ is
a qubit and $\pi(\alpha)=n(\alpha)$,
(2)
when $\alpha=(\alpha_1,\alpha_0)$
where $\alpha_0,\alpha_1$ are both
qubits and
$\pi(\alpha)=n(\alpha_0)\ol{n}(\alpha_1)$.

\section{Labelling and Summing Unitaries}
\label{sec-label-sum-uni}
In this section, we will
review a technique that we like to
call
``labelling
and summing unitaries".
This technique is also
frequently used in quantum
computation circuits, and
will be used later on in this paper,
inside more complicated circuits.

Let
$\alpha$ be
a qu(d)it for some $d\geq 2$.
Let $U$
be a
$d$ dimensional
unitary matrix.
First we will
consider the case
that $\beta$
is a qubit.

One finds that

\beq
\begin{array}{c}
U(\alpha)^{n(\beta)}\\
\end{array}
\begin{array}{c}
\ket{\psi}_\alpha
\\
H(\beta)\ket{0}_\beta
\end{array}
=
\frac{1}{\sqrt{2}}
\left(
\begin{array}{c}
\ket{\psi}_\alpha
\\
\ket{0}_\beta
\end{array}
+
\begin{array}{c}
U\ket{\psi}_\alpha
\\
\ket{1}_\beta
\end{array}
\right)
\;
\label{eq-label-unitaries}
\eeq
and

\beq
\begin{array}{c}
\\
H(\beta)
\end{array}
\begin{array}{c}
U(\alpha)^{n(\beta)}
\\
\end{array}
\begin{array}{c}
\ket{\psi}_\alpha
\\
H(\beta)\ket{0}_\beta
\end{array}
=
\begin{array}{c}
\left(\frac{1+U}{2}\right)
\ket{\psi}_\alpha
\\
\ket{0}_\beta
\end{array}
+
\begin{array}{c}
\left(\frac{1-U}{2}\right)
\ket{\psi}_\alpha
\\
\ket{1}_\beta
\end{array}
\;.
\label{eq-sum-unitaries}
\eeq
One can say
that the unitaries
$1$ and $U$ are labelled
by Eq.(\ref{eq-label-unitaries}),
and they are
summed, in the coefficient
of $\ket{0}_\beta$,
in Eq.(\ref{eq-sum-unitaries}).

So far
we have considered
$\alpha$ to be
a qu(d)it for
arbitrary $d\geq 2$,
but we have
restricted $\beta$
to be a qubit.
Let's next
consider a $\beta$
which has more than 2
independent states.
For concreteness, suppose
$\beta$ is a qu(3)it.
Let $T^{(3)}$
be a 3 dimensional unitary matrix
that satisfies

\beq
T^{(3)}\ket{0}
=
\frac{1}{\sqrt{3}}\sum_{b=0}^{2}
\ket{b}
\;.
\eeq
Suppose $U_2,U_1,U_0$
are three 3-dimensional unitary matrices.
Then Eq.(\ref{eq-label-unitaries})
generalizes to

\beqa
\lefteqn{
\begin{array}{c}
\prod_{b=2\rarrow 0}
\left\{U_b(\alpha)^{P_b(\beta)}\right\}
\\
\end{array}
\begin{array}{c}
\ket{\psi}_\alpha
\\
T^{(3)}(\beta)\ket{0}_\beta
\end{array}
= }
\\
&=&
\frac{1}{\sqrt{3}}
\left(
\begin{array}{c}
U_2\ket{\psi}_\alpha
\\
\ket{2}_\beta
\end{array}
+
\begin{array}{c}
U_1\ket{\psi}_\alpha
\\
\ket{1}_\beta
\end{array}
+
\begin{array}{c}
U_0\ket{\psi}_\alpha
\\
\ket{0}_\beta
\end{array}
\right)
\;,
\eeqa
and
Eq.(\ref{eq-sum-unitaries})
generalizes to

\beqa
\begin{array}{c}
\\
T^{(3)\dagger}(\beta)
\end{array}
\lefteqn{
\begin{array}{c}
\prod_{b=2\rarrow 0}
\left\{U_b(\alpha)^{P_b(\beta)}\right\}
\\
\end{array}
\begin{array}{c}
\ket{\psi}_\alpha
\\
T^{(3)}(\beta)\ket{0}_\beta
\end{array}
= }
\\
&=&
\underbrace{
\begin{array}{c}
\frac{1}{3}
(U_2
+
U_1
+
U_0)\ket{\psi}_\alpha
\\
\ket{0}_\beta
\end{array}
}
_{=
\begin{array}{c}
z_0 \ket{\psi_0}_\alpha
\\
\ket{0}_\beta
\end{array}
}
+
\sum_{b=2,1}
\begin{array}{c}
z_b \ket{\psi_b}_\alpha
\\
\ket{b}_\beta
\end{array}
\;,
\eeqa
where
$\sum_{b=0}^2|z_b|^2=1$
and $\av{\psi_b|\psi_b}=1$
for all $b$.
Note that
one or more of the $U_b$ can
be equal to 1.

\section{The operators $V^{(\lam)}_0$ and
$V^{(\lam)}_1$}
\label{sec-v-ops}

In the preceding Sec.\ref{sec-label-sum-uni},
we used operators
$H(\beta)$
and $T^{(3)}(\beta)$
to ``label" a set of unitary matrices
$\{1,U\}$ and
$\{U_2,U_1,U_0\}$, respectively.
In this section,
we will define
new operators
$V^{(\lam)}_0$ and
$V^{(\lam)}_1$,
where $\lam=1,2,3,\ldots$,
that will be used
in later circuits
of this paper
in a similar role,
as ``label producers" or ``labellers"
of a set of unitary matrices.

Throughout this section,
let $\lam\in\{1,2,3, \ldots\}$
and $m\in\{0,1\}$.

For $\lam=4$ and $m=0,1$,
define

\beq
V^{(4)}_m =
\begin{array}{c}
\Qcircuit @C=1em @R=1em @!R{
&\ogate
&\ogate
&\ogate
&\gate{R_y^0}
&\qw
\\
&\ogate
&\ogate
&\gate{R_y^1}\qwx[-1]
&\qw
&\qw
\\
&\ogate
&\gate{R_y^2}\qwx[-2]
&\qw
&\qw
&\qw
\\
&\gate{R_y^3}\qwx[-3]
&\qw
&\qw
&\qw
&\qw
}
\end{array}
\;,
\label{eq-vtm4}
\eeq
where

\beq
R_y^r = \exp(
-i\sigma_Y \theta_r)
\;
\eeq
for row $r=0,1,2,3$.
The angles
$\{\theta_r: r=0,1,2,3\}$
for both $m=0$ and $m=1$
will be specified later on.
$V^{(\lam)}_m$
for $\lam$ other
than 4 is defined by analogy
to Eq.(\ref{eq-vtm4}).

Below,
we will use the shorthand
notations
\beq
C_r = \cos(\theta_r)
\;,\;\;
S_r = \sin(\theta_r)
\;
\eeq
and
\beq
\ket{\{1,0^{\lam-1}\}}=
\ket{1 0^{\lam-1}}
+
\ket{0 1 0^{\lam-2}}
+
\ket{0^2 1 0^{\lam-3}}
+
\ldots
+
\ket{0^{\lam-1} 1}
\;.
\eeq

\begin{claim}\label{cl-cs-values-4}
If

\beq
\begin{array}{cccc}
S_0=\sqrt{\frac{1}{5}}
&
S_1=\sqrt{\frac{1}{4}}
&
S_2=\sqrt{\frac{1}{3}}
&
S_3=\sqrt{\frac{1}{2}}
\\
C_0=\sqrt{\frac{4}{5}}
&
C_1=\sqrt{\frac{3}{4}}
&
C_2=\sqrt{\frac{2}{3}}
&
C_3=\sqrt{\frac{1}{2}}
\end{array}
\;,
\label{eq-cs-values-4}
\eeq
then
$V^{(4)}_1$ maps

\beq
V^{(4)}_1:
\ket{0^4}\mapsto \frac{1}{\sqrt{5}}\left[
\ket{0^4}+
\ket{\{1,0^{3}\}}
\right]
\;.
\label{eq-maps-vt1}
\eeq
From Eq.(\ref{eq-maps-vt1}) it follows that
for $b^4\in Bool^4$,

\beq
\av{ b^4 | V^{(4)}_1 |0^4}=
\frac{1}{\sqrt{5}}
\left\{
\begin{array}{r}
\olb_3 \olb_2\olb_1 b_0
\\
+ \olb_3 \olb_2 b_1 \olb_0
\\
+ \olb_3 b_2\olb_1 \olb_0
\\
+ b_3 \olb_2\olb_1 \olb_0
\\
+ \olb_3 \olb_2\olb_1 \olb_0
\end{array}
\right.
\;.
\label{eq-v4m-matrix-el}
\eeq
\end{claim}
\proof
One has that

\beqa
A(b^4)&=&
\left[\;\;\;\;
\begin{array}{c}
\Qcircuit @C=1em @R=1em @!R{
\bra{b_0}\;\;\;\;\;
&\ogate
&\ogate
&\ogate
&\emptygate
&\qw
\;\;\;\;\;\ket{0}_{\beta_0}
\\
\bra{b_1}\;\;\;\;\;
&\ogate
&\ogate
&\emptygate\qwx[-1]
&\qw
&\qw
\;\;\;\;\;\ket{0}_{\beta_1}
\\\bra{b_2}\;\;\;\;\;
&\ogate
&\emptygate\qwx[-2]
&\qw
&\qw
&\qw
\;\;\;\;\;\ket{0}_{\beta_2}
\\\bra{b_3}\;\;\;\;\;
&\emptygate\qwx[-3]
&\qw
&\qw
&\qw
&\qw
\;\;\;\;\;\ket{0}_{\beta_3}
}
\end{array}
\;\;\;\;
\right]
\\
&=&
\begin{array}{l}
\bra{b_0}
\\
\bra{b_1}
\\
\bra{b_2}
\\
\bra{b_3}
\end{array}
\left[
\begin{array}{c}
\exp\{-i\sigma_Y(\beta_0)\theta_0\}
\\
\exp\{-i\sigma_Y(\beta_1)\theta_1 P_0(\beta_0)\}
\\
\exp\{-i\sigma_Y(\beta_2)\theta_2
P_{0}(\beta_1)P_{0}(\beta_0)
\}
\\
\exp\{-i\sigma_Y(\beta_3)\theta_3
P_{0}(\beta_2)P_{0}(\beta_1)P_{0}(\beta_0)
\}
\end{array}
\right]
\ket{0^4}
\;.
\eeqa

It's easy to convince oneself that
the only non-vanishing matrix elements are those
for which $b^4$
has either (1) all 4 components
equal to 0, or (2)
a single component equal to 1 and  the other 3 components equal to 0.
Evaluating each of these possibilities separately, one finds

\beq
A(b^4)=
\left\{
\begin{array}{r}
S_0\; \olb_3 \olb_2\olb_1 b_0
\\
+S_1 C_0\; \olb_3 \olb_2 b_1 \olb_0
\\
+S_2 C_1 C_0 \; \olb_3 b_2\olb_1 \olb_0
\\
+S_3 C_2 C_1 C_0 \;  b_3 \olb_2\olb_1 \olb_0
\\
+C_3 C_2 C_1 C_0 \; \olb_3 \olb_2\olb_1 \olb_0
\end{array}
\right.
\;.
\label{eq-a-a4}
\eeq
Now one can
plug
into Eq.(\ref{eq-a-a4})
the values of $C_r$ and $S_r$
given in the premise of our
claim to show that the conclusion
of our claim holds.
\qed

\begin{claim}\label{cl-cs-values-4-neg}
If
$C_r$ and $S_r$ for $r=3,2,1,0$
have the values given by
Eqs.(\ref{eq-cs-values-4}),
then
$V^{(4)}_1$ maps

\beq
V^{(4)}_1:
\begin{array}{c}
\ket{0}
\\
\ket{0}
\\
\ket{0}
\\
\ket{1}
\end{array}
\mapsto
\frac{1}{\sqrt{5}}\left[
-\ket{0^4}+
\ket{\{1,0^{3}\}}
\right]
\;.
\label{eq-maps-vt1-neg}
\eeq
From Eq.(\ref{eq-maps-vt1-neg}) it follows that
for $b^4\in Bool^4$,

\beq
\bra{b^4 } V^{(4)}_1
\begin{array}{c}
\ket{0}
\\
\ket{0}
\\
\ket{0}
\\
\ket{1}
\end{array}=
\frac{1}{\sqrt{5}}
\left\{
\begin{array}{r}
\olb_3 \olb_2\olb_1 b_0
\\
+ \olb_3 \olb_2 b_1 \olb_0
\\
+ \olb_3 b_2\olb_1 \olb_0
\\
+ b_3 \olb_2\olb_1 \olb_0
\\
- \olb_3 \olb_2\olb_1 \olb_0
\end{array}
\right.
\;.
\label{eq-v4m-matrix-el-neg}
\eeq
\end{claim}
\proof

Eq.(\ref{eq-a-a4})
is true in this case,
but only if
we replace $C_3\rarrow -S_3=-\frac{1}{\sqrt{2}}$
and $S_3\rarrow C_3=\frac{1}{\sqrt{2}}$.
\qed

\begin{claim}\label{cl-cs-values-4-zero}
If

\beq
\begin{array}{cccc}
S_0=\sqrt{\frac{1}{4}}
&
S_1=\sqrt{\frac{1}{3}}
&
S_2=\sqrt{\frac{1}{2}}
&
S_3=1
\\
C_0=\sqrt{\frac{3}{4}}
&
C_1=\sqrt{\frac{2}{3}}
&
C_2=\sqrt{\frac{1}{2}}
&
C_3=0
\end{array}
\;,
\label{eq-cs-values-4-zero}
\eeq
then $V^{(4)}_0$ maps
\beq
V^{(4)}_0:
\ket{0^4}\mapsto
\frac{1}{\sqrt{4}}
\ket{\{1,0^{3}\}}
\;.
\label{eq-maps-vt0}
\eeq
From Eq.(\ref{eq-maps-vt0}) it follows that
for $b^4\in Bool^4$,

\beq
\av{ b^4 | V^{(4)}_0 |0^4}=
\frac{1}{\sqrt{4}}
\left\{
\begin{array}{r}
\olb_3 \olb_2\olb_1 b_0
\\
+ \olb_3 \olb_2 b_1 \olb_0
\\
+ \olb_3 b_2\olb_1 \olb_0
\\
+ b_3 \olb_2\olb_1 \olb_0
\end{array}
\right.
\;.
\label{eq-v4m-matrix-el-zero}
\eeq
Note that
$C_3=0,S_3=1$
means $\theta_3 = \pi/2$, and
$e^{-i\sigma_Y \theta_3}=-i\sigma_Y$.
\end{claim}
\proof

Plug
into Eq.(\ref{eq-a-a4})
the values of $C_r$ and $S_r$
given in the premise of our
claim to show that the conclusion
of our claim holds.
\qed

\begin{claim}
$V^{(\lam)}_m$
for $\lam$ other
than 4
satisfies
claims analogous to Claims \ref{cl-cs-values-4},
\ref{cl-cs-values-4-neg},
and
\ref{cl-cs-values-4-zero}.
\end{claim}
\proof
Obvious.
\qed

\section{Targeting
Two Hypotheses}
\label{sec-cats}

In this section, we will
describe a simple trick that
can sometimes be used
when applying Grover's original
algorithm or some variant thereof
like AFGA, as long as it
drives a
starting state $\ket{s}$
to a target state $\ket{t}$.
Sometimes it is possible
to arrange things
so that the target
state is a
superposition
$a_0\ket{0}+a_1\ket{1}$
of two orthonormal states $\ket{0}$ and $\ket{1}$,
so that if we know $a_0$, we can
infer $a_1$,
a type of hypothesis testing
with 2 hypotheses.
If the target state
were just proportional
to say $\ket{0}$,
then its component
along $\ket{0}$
would be 1 after normalization
so one wouldn't be able
to do any type of amplitude inference.

Suppose $z_0, z_1$ are
complex numbers
and $\ket{\chi}$
is an unnormalized state
such that

\beq
|z_0|^2 + |z_1|^2 + \av{\chi|\chi} =1
\;.
\eeq
Define

\beq
p = |z_0|^2 + |z_1|^2
\;,\;\; q= 1-p
\;,
\eeq
and

\beq
\hat{z}_0 = \frac{z_0}{\sqrt{p}}\;,\;\;
\hat{z}_1 = \frac{z_1}{\sqrt{p}}
\;.
\eeq

Assume the
states $\{\ket{\psi_j}_\mu\}_{j=0,1}$
are orthonormal,
the states
$\{\ket{j}_\nu\}_{j=0,1}$
are orthonormal,
and the states
$\{\ket{b}_\omega \}_{b=0,1}$
are orthonormal.

We wish to do AFGA with
the following starting state
$\ket{s}_{\mu, \nu,\omega}$
 and target state
$\ket{t}_{\mu, \nu,\omega}$:

\beq
\ket{s}_{\mu,\nu,\omega}=
\begin{array}{c}
z_0 \ket{\psi_0}_\mu
\\
\ket{0}_\nu
\\
\ket{0}_\omega
\end{array}
+
\begin{array}{c}
z_1 \ket{\psi_1}_\mu
\\
\ket{1}_\nu
\\
\ket{0}_\omega
\end{array}
+
\begin{array}{c}
\ket{\chi}_{\mu, \nu}
\\
\ket{1}_\omega
\end{array}
\;
\eeq
and

\beq
\ket{t}_{\mu, \nu,\omega}=
\begin{array}{c}
\hat{z}_0\ket{\psi_0}_\mu
\\
\ket{0}_\nu
\\
\ket{0}_\omega
\end{array}
+
\begin{array}{c}
\hat{z}_1\ket{\psi_1}_\mu
\\
\ket{1}_\nu
\\
\ket{0}_\omega
\end{array}
\;.
\label{eq-t-full}
\eeq
We will refer
to $\ket{0}_\nu$
as the null hypothesis
state, and
to $\ket{1}_\nu$
as the alternative or rival
hypothesis state.

From the previous definitions,
one finds

\beq
\begin{array}{r}
\left[\ket{t}\bra{t}\right]_{\mu,\nu,\omega} \ket{s}_{\mu,\nu,\omega}=
\sqrt{p} \;\;\ket{t}_{\mu,\nu,\omega}
\\
\left[\ket{0}\bra{0}\right]_{\omega} \ket{s}_{\mu,\nu,\omega}=
\sqrt{p} \;\;\ket{t}_{\mu,\nu,\omega}
\end{array}
\;
\eeq
and

\beq
\begin{array}{r}
\left[\ket{t}\bra{t}\right]_{\mu,\nu,\omega} \ket{t}_{\mu,\nu,\omega}= \;\;\ket{t}_{\mu,\nu,\omega}
\\
\left[\ket{0}\bra{0}\right]_{\omega} \ket{t}_{\mu,\nu,\omega}=
 \;\;\ket{t}_{\mu,\nu,\omega}
\end{array}
\;.
\eeq
$\ket{t}$
only appears
in AFGA within
the projection operator
$\ket{t}\bra{t}$,
and this projection operator
always acts solely on the space spanned by
$\ket{t}$ and $\ket{s}$.
But $\ket{t}\bra{t}$
and
$\ket{0}\bra{0}_\omega$
act identically on that space.
Hence, for the purposes
of AFGA, we can
replace
$\ket{t}\bra{t}$
by
$\ket{0}\bra{0}_\omega$.
We will call
$\ket{0}_\omega$
 the ``sufficient" target state
 to distinguish it from the
full target state $\ket{t}_{\mu, \nu,\omega}$.

Recall that AFGA converges in order
$\frac{1}{|\av{t|s}|}$ steps. From the
definitions of $\ket{s}$ and
$\ket{t}$, one finds

\beq
\av{t|s} = \sqrt{p}
\;.
\label{eq-st-sqrt-p}
\eeq

Once system $(\mu,\nu,\omega)$
has been driven
to the target state $\ket{t}_{\mu,\nu,\omega}$,
one can measure
the subsystem $\nu$
while ignoring the subsystem $(\mu,\omega)$.
If we
do so,
the outcome
of the measurements of
$\nu$ can be
predicted from the
partial density matrix:

\beq
\tr_{\mu,\omega} \left\{\ket{t}\bra{t}_{\mu,\nu,\omega}\right\}=
P(0) \ket{0}\bra{0}_\nu
+
P(1) \ket{1}\bra{1}_\nu
\;,
\eeq
where

\beq
P(0) = |\hat{z}_0|^2
\;,\;\;
P(1) = |\hat{z}_1|^2
\;.
\eeq
Hence

\beq
|z_1|^2 = \frac{P(1)}{P(0)} |z_0|^2
\;.
\label{eq-z1-z0}
\eeq
We see that
$|z_1|$ and $|z_0|$
are proportional to each other,
with a proportionality factor
that can be
calculated by measuring the
subsystem $\nu$
multiple times.
If we know
$|z_0|$, we can use Eq.(\ref{eq-z1-z0}) to find $|z_1|$.
More generally, if $|z_j|^2=f_j(\theta)$
for $j=0,1$, and  the functions $f_j()$
are known but the parameter $\theta$
isn't, we can solve
$f_1(\theta)/f_0(\theta)=P(1)/P(0)$
for $\theta$.

Eq.(\ref{eq-z1-z0}) only relates the magnitudes of
$z_0$ and $z_1$. One can also measure
the relative phase between
$z_0$ and $z_1$ as follows.
Let $z_0(z_1)^*=|z_0 z_1|e^{i\theta}$.
Before taking the final measurement of $\nu$, apply a unitary transformation that maps
$\ket{t}$ given by Eq.(\ref{eq-t-full}) to
$\ket{t'}$ given by

\beq
\ket{t'}_{\mu, \nu,\omega}=
\left(\frac{\hat{z}_0 + \hat{z}_1}{\sqrt{2}}\right)
\begin{array}{r}
\ket{\psi_0}_\mu
\\
\ket{0}_\nu
\\
\ket{0}_\omega
\end{array}
+
\left(\frac{\hat{z}_0 - \hat{z}_1}{\sqrt{2}}\right)
\begin{array}{r}
\ket{\psi_1}_\mu
\\
\ket{1}_\nu
\\
\ket{0}_\omega
\end{array}
\;.
\eeq
Then do as before, measure
$\nu$ in the $\{\ket{0},\ket{1}\}$
basis while ignoring
$(\mu,\omega)$.
If we
do so,
the outcome
of the measurements of
$\nu$ can be
predicted from the
partial density matrix:

\beq
\tr_{\mu,\omega} \left\{\ket{t'}\bra{t'}_{\mu,\nu,\omega}
\right\}=
P(+) \ket{0}\bra{0}_\nu
+
P(-) \ket{1}\bra{1}_\nu
\;,
\eeq
where

\beqa
P(\pm) &=&
\frac{1}{2}
|\hat{z}_0\pm \hat{z}_1|^2
\\
&=&
\frac{1}{2}
\left[
P(0) + P(1) \pm 2P(0)P(1)\cos \theta
\right]
\;.
\eeqa
Hence,

\beq
\cos \theta =
\frac{P(+)-P(-)}{2P(0)P(1)}
\;.
\label{eq-z01-cos-theta}
\eeq

\section{Blind Targeting}\label{sec-blind}
At first sight,
it seems that Grover-like algorithms and AFGA
in particular require knowledge of
$|\av{t|s}|$. In this section, we will
describe a technique for bypassing that
onerous requirement.

For concreteness, we will
assume in our discussion below that we are using AFGA
and that we are targeting two hypotheses, but the idea of this
technique could be carried over to
other Grover-like algorithms in a
fairly obvious way.

According to Eq.(\ref{eq-st-sqrt-p}), when targeting two hypotheses, $|\av{t|s}| = \sqrt{p}$.
Suppose we guess-timate $p$, and use that
estimate and the AFGA formulas of Ref.\cite{afga}
to calculate the various
rotation angles $\alpha_j$ for $j=0,1,\ldots,N_{Gro}-1$,
where $N_{Gro}$ is the number of Grover steps.
Suppose $N_{Gro}$ is large enough.
Then, in the unlikely event that our estimate of $p$
is perfect, $\hat{s}_j$ will converge to
$\hat{t}$ as $j\rarrow N_{Gro}-1$. On the other
hand, if our estimate of $p$
is not perfect but not too bad either,
we expect that as $j\rarrow N_{Gro}-1$,
the point $\hat{s}_j$ will reach a steady
state in which, as $j$ increases, $\hat{s}_j$ rotates in a
small circle in the neighborhood of $\hat{t}$. After steady state is reached,
all functions of $\hat{s}_j$ will vary periodically with $j$.

Suppose we do AFGA
with $p$ fixed and with
$N_{Gro}=(N_{Gro})_0 + r$ Grover steps
where $r=0,1,\ldots N_{tail}-1$.
 Call each $r$
 a ``tail run", so
 $p$ is the same
 for all $N_{tail}$ tail runs, but $N_{Gro}$
 varies for different tail runs.
Suppose that steady state
has already been reached after $(N_{Gro})_0$
steps.
For any quantity $Q_r$
where $r=0,1,\ldots N_{tail}-1$,
let  $\avss{Q}$
denote the outcome of passing the
$N_{tail}$ values of $Q_r$
through a low pass filter that
takes out the AC components and leaves
only the DC part.
For example,
$\avss{Q}$ might
equal $\sum_r Q_r/N_{tail}$
or $[\max_r{Q_r} + \min_r{Q_r}]/2$.
By applying the SEO of tail run
$r$ to a quantum computer several times, each time ending with
a measurement of the quantum computer, we can obtain
values $P_r(0)$
and $P_r(1)$ of $P(0)$ and $P(1)$
for tail run $r$.
Then we can find $\avss{\sqrt{P(1)/P(0)}} =\avss{|z_1|/|z_0|}$.
But we also expect to know
$|z_0|$, so we can use
$\avss{|z_1|/|z_0|}|z_0|$
as an estimate of $|z_1|$.
This estimate of
$|z_1|$ and the known value of
$|z_0|$ yield a new estimate of
$p=|z_1|^2 + |z_0|^2$, one that is much better than the first estimate we used. We can repeat the previous steps using  this new estimate of
$p$.
Every time we repeat this process, we get a new estimate of
$p$ that is better than
our previous estimate.
Call a ``trial" each time we repeat
the process of $N_{tail}$ tail runs.
$p$
is fixed
during a trial, but $p$ varies
from trial to trial.

Appendix \ref{app-blind}
describes a numerical experiment
that
we performed. The experiment provides
some evidence that our blind targeting
technique
behaves as we say it does
when used in conjunction with AFGA.

\section{Quantum Circuit For
Calculating $|\av{c^n|\pi_{Sym_n}|\psi}|^2$}

In this section, we will
give the main quantum
circuit of this paper,
one that can be used
to calculate
$|\av{c^n|\pi_{Sym_n}|\psi}|^2$
for some predetermined point $c^n\in\{0..d-1\}^n$
and
state $\ket{\psi}_{\alpha^n}$,
where $width(\alpha^n)=d^n$.
Actually, in this paper,
we will give two alternative
methods for calculating
$|\av{c^n|\pi_{Sym_n}|\psi}|^2$.
The method presented in this section
will be called Method A. Appendix \ref{app-method-b} presents an alternative method that will
be called Method B.

We will assume that
we know how to compile
$\ket{\psi}_{\alpha^n}$
(i.e., that
we can construct it starting
from $\ket{0^n}_{\alpha^n}$
using a sequence of
elementary operations.
Elementary operations are
operations that act on a few (usually 1,2 or 3)
qubits at a time,
such as qubit rotations
and CNOTS.)
Multiplexor techniques for doing such compilations
are discussed in Ref.\cite{tuc-multiplexor}.
If $n$
is very large,
our algorithm will
be useless unless
such a compilation
is of polynomial efficiency,
meaning that
its number of elementary
operations grows as poly($n$).

For concreteness, henceforth
we will use $n=4$
in this section,
but it will be obvious
how to draw
an analogous
circuit
for arbitrary $n$.

For $r=4,3,2,1$, define

\beq
Q^{(r)}(c^4)
=
|\bra{c^4}
\pi_{Sym_r}(\alpha_{\leq r-1})\ket{\psi}_{\alpha^4}|^2
\;
\eeq
where
$\alpha_{\leq r-1}=(\alpha_{r-1},\ldots, \alpha_1,\alpha_0)$.
For instance,

\beq
Q^{(1)}(c^4)= |\av{c^4|\psi}|^2
\;
\eeq
and

\beq
Q^{(2)}(c^4)= |\bra{c^4}
\pi_{Sym_2}(\alpha_1,\alpha_0)
\ket{\psi}_{\alpha^4}|^2
\;.
\eeq

We want
all
horizontal lines
in Fig.\ref{fig-sym-ckt}
to represent qubits,
except for the $\alpha_j$ lines
which should represent qu(d)its.
Let $\alpha = \alpha^4$
and
$\beta=
(\beta^1_{;0},\beta^2_{;1},\beta^3_{;2})$.

Define
\beq
T(\alpha, \beta^1_{;0})
=
V_1^{(1)\dagger}(\beta^1_{;0})
\left[
\begin{array}{c}
\Qcircuit @C=1em @R=1em @!R{
&\lamgate&\qw&\scriptstyle{\alpha_0}
\\
&\veegate&\qw&\scriptstyle{\alpha_1}
\\
&\qw&\qw&\scriptstyle{\alpha_2}
\\
&\qw&\qw&\scriptstyle{\alpha_3}
\\
&\dotgate\qwx[-4]&\qw&\scriptstyle{\beta_{0;0}}
}
\end{array}
\right]
V_1^{(1)}(\beta^1_{;0})
\;,
\eeq

\beq
T(\alpha, \beta^2_{;1})=
V_1^{(2)\dagger}(\beta^2_{;1})
\left[
\begin{array}{c}
\Qcircuit @C=1em @R=1em @!R{
&\lamgate&\qw&\qw&\scriptstyle{\alpha_0}
\\
&\qw&\lamgate&\qw&\scriptstyle{\alpha_1}
\\
&\veegate&\veegate&\qw&\scriptstyle{\alpha_2}
\\
&\qw&\qw&\qw&\scriptstyle{\alpha_3}
\\
&\qw&\dotgate\qwx[-3]&\qw&\scriptstyle{\beta_{0;1}}
\\
&\dotgate\qwx[-5]&\qw&\qw&\scriptstyle{\beta_{1;1}}
}
\end{array}
\right]
V_1^{(2)}(\beta^2_{;1})
\;,
\eeq

\beq
T(\alpha, \beta^3_{;2})=
V_1^{(3)\dagger}(\beta^3_{;2})
\left[
\begin{array}{c}
\Qcircuit @C=1em @R=1em @!R{
&\lamgate&\qw&\qw&\qw&\scriptstyle{\alpha_0}
\\
&\qw&\lamgate&\qw&\qw&\scriptstyle{\alpha_1}
\\
&\qw&\qw&\lamgate&\qw&\scriptstyle{\alpha_2}
\\
&\veegate&\veegate&\veegate&\qw&\scriptstyle{\alpha_3}
\\
&\qw&\qw&\dotgate\qwx[-2]&\qw&\scriptstyle{\beta_{0;2}}
\\
&\qw&\dotgate\qwx[-4]&\qw&\qw&\scriptstyle{\beta_{1;2}}
\\
&\dotgate\qwx[-6]&\qw&\qw&\qw&\scriptstyle{\beta_{2;2}}
}
\end{array}
\right]
V_1^{(3)}(\beta^3_{;2})
\;,
\eeq

\beq
T(\alpha,\beta) =\prod_{\ell=2\rarrow 0}T(\alpha, \beta_{;\ell})
\;,
\eeq

\beq
\pi(\alpha)=\prod_{j=0}^3 P_{c_j}(\alpha_j)
\;
\eeq
and

\beq
\pi(\beta)=
\left\{
\begin{array}{l}
P_0(\beta_{0;0})
\\
P_0(\beta_{0;1})
P_0(\beta_{1;1})
\\
P_0(\beta_{0;2})
P_0(\beta_{1;2})
P_0(\beta_{2;2})
\end{array}
\right.
\;.
\label{eq-def-pi-beta}
\eeq

\subsection{Method A}\label{sec-method-a}
\begin{figure}[h]
    \begin{center}
    \epsfig{file=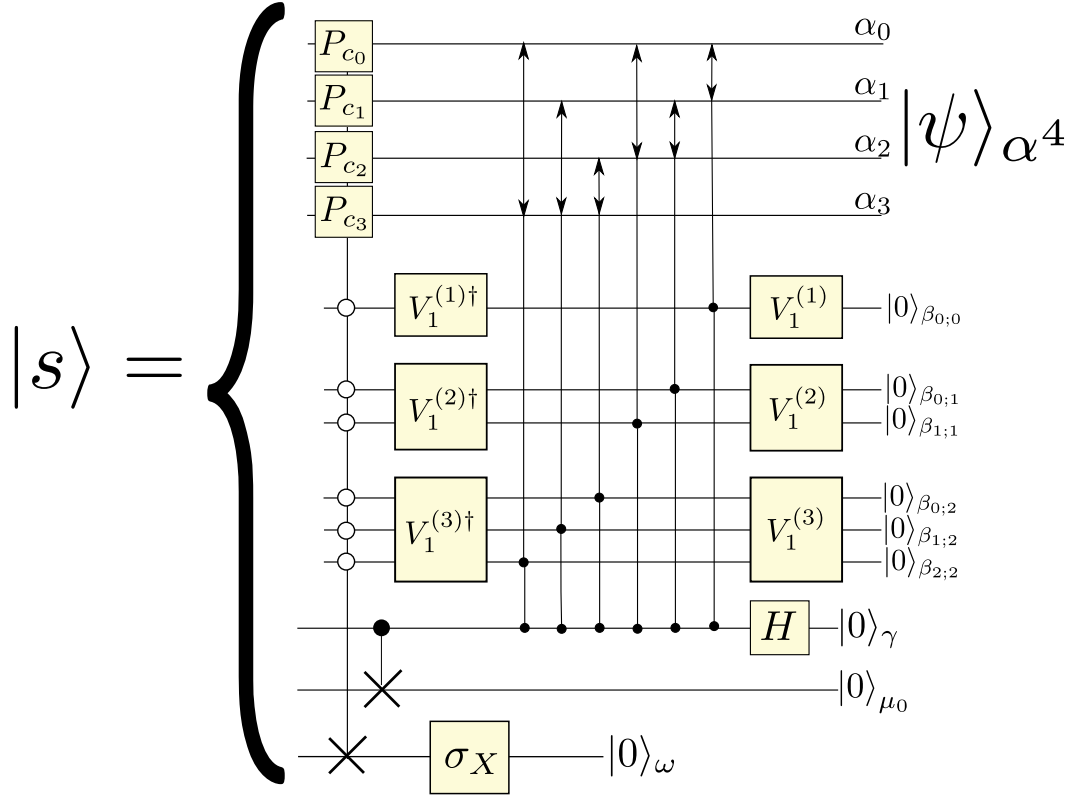, width=5.5in}
    \caption{Method A circuit for
    generating $\ket{s}$
    used in AFGA to calculate
    $|\av{c^4|\pi_{Sym_4}|\psi}|^2$
    }
    \label{fig-sym-ckt}
    \end{center}
\end{figure}

Method A
for
calculating
$Q^{(4)}(c^4)$
consists of applying the algorithm
AFGA of Ref.\cite{afga}
in the way that was described in
Sec.\ref{sec-cats},
using the techniques
of targeting two hypotheses
and blind targeting.
As in Sec.\ref{sec-cats},
when we apply AFGA in this section,
we will use a sufficient target $\ket{0}_\omega$.
All that remains for
us to do to
fully specify our
circuit for calculating
$Q^{(4)}(c^4)$
is to give a circuit for
generating $\ket{s}$.

A circuit for generating
$\ket{s}$ is given by
Fig. \ref{fig-sym-ckt}.
Fig.\ref{fig-sym-ckt}
is equivalent to saying that

\beq
\ket{s}_{\mu,\nu,\omega}=
\sigma_X(\omega)^{
\pi(\beta)
\pi(\alpha)}
\frac{1}{\sqrt{2}}
\left[
\begin{array}{l}
T(\alpha,\beta)
\begin{array}{r}
\ket{\psi}_{\alpha^4}
\\
\ket{0^6}_\beta
\end{array}
\\
\ket{1}_\gamma
\\
\ket{1}_{\mu_0}
\\
\ket{1}_\omega
\end{array}
+
\begin{array}{l}
\ket{\psi}_{\alpha^4}
\\
\ket{0^6}_\beta
\\
\ket{0}_\gamma
\\
\ket{0}_{\mu_0}
\\
\ket{1}_\omega
\end{array}
\right]
\;.
\eeq

\begin{claim}

\beq
\ket{s}_{\mu,\nu,\omega}=
\begin{array}{c}
z_1 \ket{\psi_1}_\mu
\\
\ket{1}_\nu
\\
\ket{0}_\omega
\end{array}
+
\begin{array}{c}
z_0 \ket{\psi_0}_\mu
\\
\ket{0}_\nu
\\
\ket{0}_\omega
\end{array}
+
\begin{array}{c}
\ket{\chi}_{\mu,\nu}
\\
\ket{1}_\omega
\end{array}
\;,
\label{eq-start-with-chi}
\eeq
for some unnormalized state
$\ket{\chi}_{\mu,\nu}$,
where

\beq
\begin{array}{|c|c|}
\hline
\ket{\psi_1}_\mu=
\begin{array}{r}
\ket{c^4}_\alpha
\\
\ket{1}_{\mu_0}
\end{array}
&
\ket{\psi_0}_\mu=
\begin{array}{r}
\ket{c^4}_\alpha
\\
\ket{0}_{\mu_0}
\end{array}
\\
\ket{1}_\nu=
\left[
\begin{array}{l}
\ket{0}_{\beta_{;0}}
\\
\ket{00}_{\beta_{;1}}
\\
\ket{000}_{\beta_{;2}}
\\
\ket{1}_{\gamma}
\end{array}
\right]
&
\ket{0}_\nu=
\left[
\begin{array}{l}
\ket{0}_{\beta_{;0}}
\\
\ket{00}_{\beta_{;1}}
\\
\ket{000}_{\beta_{;2}}
\\
\ket{0}_{\gamma}
\end{array}
\right]
\\
\hline
\end{array}
\;,
\eeq

\beq
z_1 = \frac{1}{\sqrt{2}}\av{c^4|\pi_{Sym_4}|\psi}
=\sqrt{\frac{Q^{(4)}(c^4)}{2}}
\;,
\eeq

\beq
z_0 = \frac{1}{\sqrt{2}}\av{c^4|\psi}
=\sqrt{\frac{Q^{(1)}(c^4)}{2}}
\;,
\eeq

\beq
\frac{|z_1|}{|z_0|} = \sqrt{\frac{P(1)}{P(0)}}
\;.
\label{eq-z1-z0-again}
\eeq
\end{claim}
\proof

Applying identity Eq.(\ref{eq-u-pi-id}) with $U=\sigma_X(\omega)$
yields:

\beqa
\ket{s}&=&
\sigma_X(\omega)^{\pi(\beta)\pi(\alpha)}
\ket{s'}
\\
&=&
\sigma_X(\omega)\pi(\beta)\pi(\alpha)
\ket{s'}
+
\begin{array}{l}
\ket{\chi}
\\
\ket{1}_\omega
\end{array}
\;.
\eeqa

Eq.(\ref{eq-z1-z0-again}) is just Eq.(\ref{eq-z1-z0}).
\qed

In case $\av{c^4|\psi}=0$,
this procedure won't yield $Q^{(4)}(c^4)$,
but it can be patched up easily.
Note that if we know how to compile
$\ket{\psi}_{\alpha^4}$
with polynomial efficiency,
then we also know how to compile
$\ket{\psi'}=swap(\alpha_0,\alpha_1)\ket{\psi}$
with polynomial efficiency.
Furthermore,

\beq
\av{c^4|\pi_{Sym_4}|\psi}=
\av{c^4|\pi_{Sym_4}|\psi'}
\;.
\eeq
If
$\av{c^4|\psi'}\neq 0$, mission accomplished.
Even if
$\av{c^4|\psi'}=0$,
as long as we can replace
$\ket{\psi}$ by some partially symmetrized
version of it, call it $\ket{\psi_S}$,
such that
$\av{c^4|\psi_S}\neq 0$,
we should be able to
apply this method
to get $Q^{(4)}(c^4)$.

\appendix
\section{Appendix: Numerical Experiment
to Test Blind Targeting with AFGA}\label{app-blind}

In this appendix, we will
describe a
numerical experiment that we conducted
to test blind targeting with AFGA.
The experiment is not a conclusive proof that
blind targeting with AFGA always
converges to the right answer, but it does
provide some evidence that
it often does.

Our algorithm for blind targeting  is
based on the following Bloch sphere picture.
We will use the notation of Ref.\cite{afga}.
Suppose we know the vector $\hat{s}_0$
but we don't know that $\hat{t}=\hat{z}$, so we don't know the initial $|\av{t|s}| =\left|\cos(\frac{1}{2}{\rm acos}(\hat{t}\cdot\hat{s}))\right|$.
Suppose we guess-timate $|\av{t|s}|$, and use that
estimate and the AFGA formulas of Ref.\cite{afga}
to calculate the unit vector $\hat{s}_j$ for $j=0,1,\ldots,N_{Gro}-1$,
where $N_{Gro}$ is the number of Grover steps.
Suppose $N_{Gro}$ is large enough.
Then, in the unlikely event that our estimate of $|\av{t|s}|$
is perfect, as $j\rarrow N_{Gro}-1$,
the point
$\hat{s}_j$ will converge to
$\hat{t}=\hat{z}$. On the other
hand, if our estimate of $|\av{t|s}|$
is not perfect but not too bad either,
we expect that as $j\rarrow N_{Gro}-1$,
the point $\hat{s}_j$ will reach a steady
state in which, as $j$ increases, $\hat{s}_j$ rotates in a
circle of constant latitude very close
to the North Pole of the Bloch sphere.

If we pass through a low pass filter the values of $\hat{s}_j$
after it reaches this steady state, we will get an estimate of the position of the North Pole.
Using that estimate $\hat{t}_{est}$ of the position of the North Pole and our assumed
knowledge of $\hat{s}$ allows us to
get a new estimate of
$|\av{t|s}|$, one that is much better than the first estimate we used. We can repeat the previous steps using  this new estimate of
$|\av{t|s}|$.
Every time we repeat this process, we get a new estimate of
$|\av{t|s}|$ that is better than
our previous estimate.

To get some numerical
evidence that this Bloch sphere picture
argument applies, we wrote a new version
of the .m files\footnote{Our .m files are written in the language of Octave. The Octave environment is a free, open source, partial clone of the MatLab environment. Octave .m files can usually be run in Matlab with zero or only minor modifications.} that were written to illustrate the AFGA algorithm of Ref.\cite{afga} and were included with
the arXiv distribution of that paper.
The arXiv distribution of the present paper includes 3 new Octave .m files: {\tt afga\_blind.m},
{\tt afga\_step.m} and {\tt afga\_rot.m}.

The files {\tt afga\_step.m} and {\tt afga\_rot.m} contain auxiliary functions called by the main file
{\tt afga\_blind.m}. These 2 files are identical to the files with the same names that were included with Ref.\cite{afga}. Hence, we will say nothing more about them
here.

The file {\tt afga\_blind.m} is a
slight expansion of the file
{\tt afga.m} that was presented and
explained in Ref.\cite{afga}.
The first 7 non-comment lines of
{\tt afga\_blind.m} instantiate
the following 7 input parameters:
\begin{itemize}
\item
{\tt g0\_degs}$=\gamma=\gamma_0$ in degrees. Used only to calculate $\hat{s}_0$, which is assumed known, not to calculate the initial $\av{t|s}$, which is assumed a priori unknown.
\item {\tt g0est\_degs} $=$ an estimate of $\gamma_0$,
in degrees. Used to get first estimate of $\av{t|s}$.
\item
{\tt del\_lam\_degs}$= \Delta \lam$ in degrees
\item
{\tt num\_steps}$=N_{Gro}=$ number of Grover steps.
\item
{\tt tail\_len} $= N_{tail}=$ tail length,
number of tail runs. Low
 pass filtering is
 applied to points $j=N_{Gro}-N_{tail},
 \ldots, N_{Gro}-3, N_{Gro}-2, N_{Gro}-1$
of each trial to get
the estimate of $\av{t|s}$ for the next trial.
\item
{\tt num\_trials} $=$ number of
trials. $\gamma_0$
remains constant during a trial,
but changes from trial to trial.
\item
{\tt plotted\_trial} $=$ trial for which program will plot the time series $\hat{s}_j$ for $j=0,1,\ldots,N_{Gro}-1$.

\end{itemize}

Each time {\tt afga\_blind.m} runs successfully, it outputs two files
called {\tt afga\_blind.txt} and
{\tt afga\_blind.svg}.

The output file {\tt afga\_blind.txt}
is a text file. Its contents are very
similar to the contents of the
file {\tt afga.txt}
that is outputted by the program
{\tt afga.m} of Ref.\cite{afga}. The contents of an
{\tt afga.txt} file are thoroughly
explained in Ref.\cite{afga}.
From that, it's very easy to understand
the meaning of the contents of an
{\tt afga\_blind.txt} file. An {\tt afga\_blind.txt} file
contains the records of {\tt num\_trials} trials instead
of just one trial like an {\tt afga.txt} file does.

The output file {\tt afga\_blind.svg}
is a picture of a plot, in .svg (scalable vector graphic) format.
.svg files can be viewed with a web browser.
They can be viewed and modified with,
for example, the free, open source software program Inkscape.
The plot in an {\tt afga\_blind.svg}
file gives the 3 components of the
unit vector $\hat{s}_j$
as a function of the Grover step $j$.
The 7 input parameters just described
are listed in a legend of the plot.

\begin{figure}[t]
    \begin{center}
    \epsfig{file=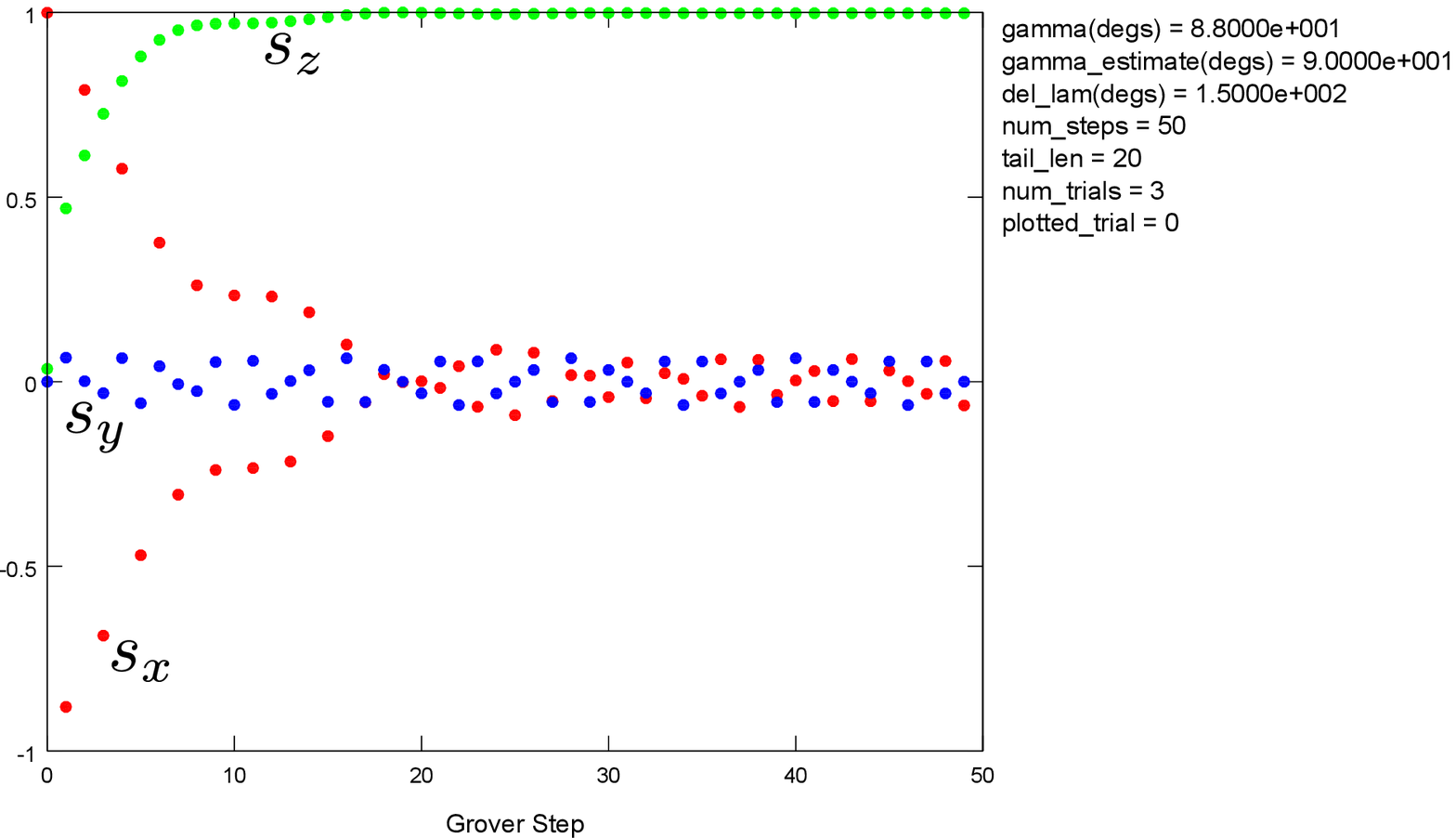, width=5in}
    \caption{
    The 3 components of the unit vector $\hat{s}_j$
    as a function of the Grover step $j$. Plot generated by {\tt afga\_blind.m} with indicated inputs.
    }
    \label{fig-afga-blind90-0}
    \end{center}
\end{figure}

\begin{figure}[b]
    \begin{center}
    \epsfig{file=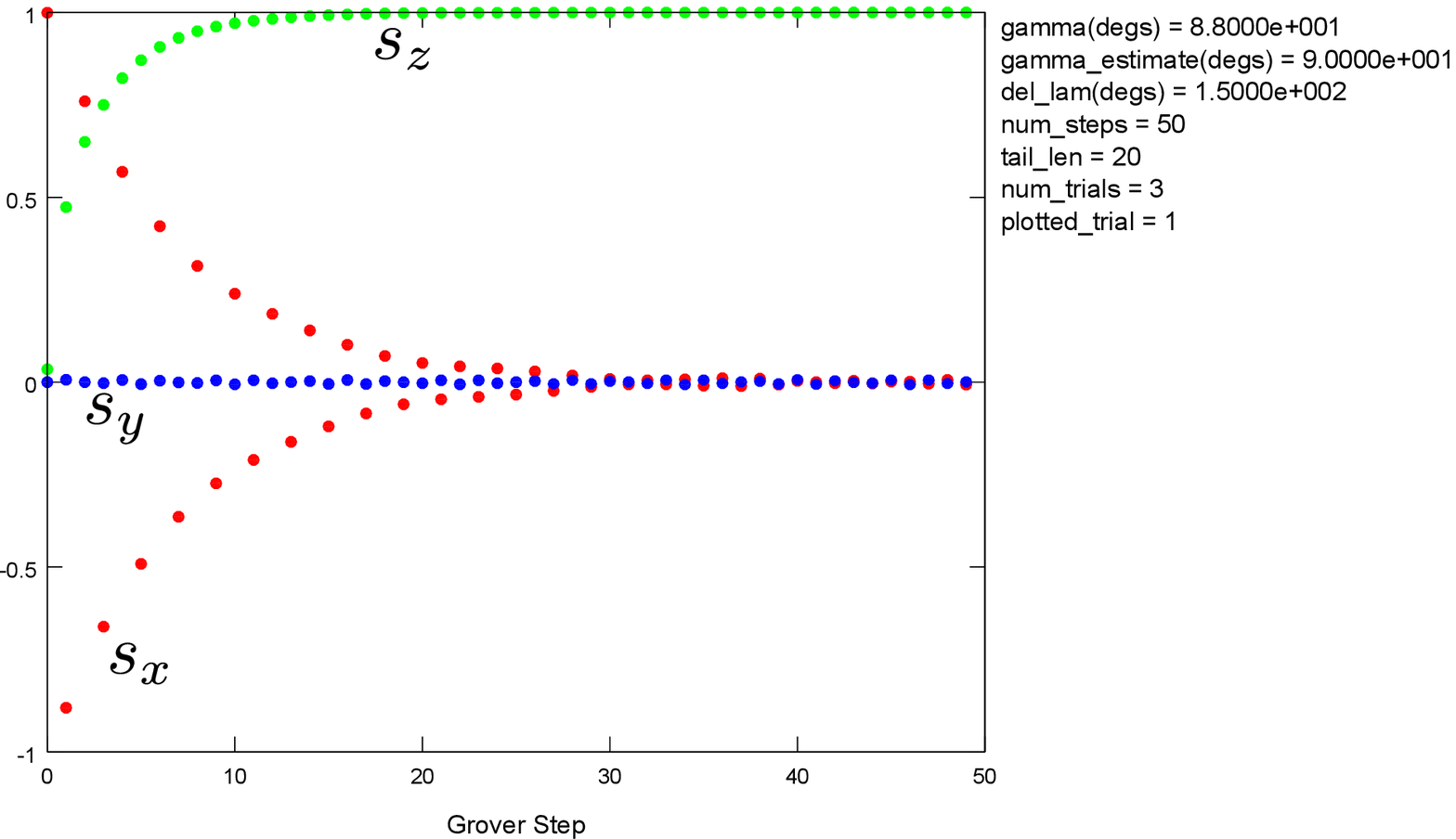, width=5in}
    \caption{
    The 3 components of the unit vector $\hat{s}_j$
    as a function of the Grover step $j$. Plot generated by {\tt afga\_blind.m} with indicated inputs.
    }
    \label{fig-afga-blind90-1}
    \end{center}
\end{figure}

\begin{figure}[t]
    \begin{center}
    \epsfig{file=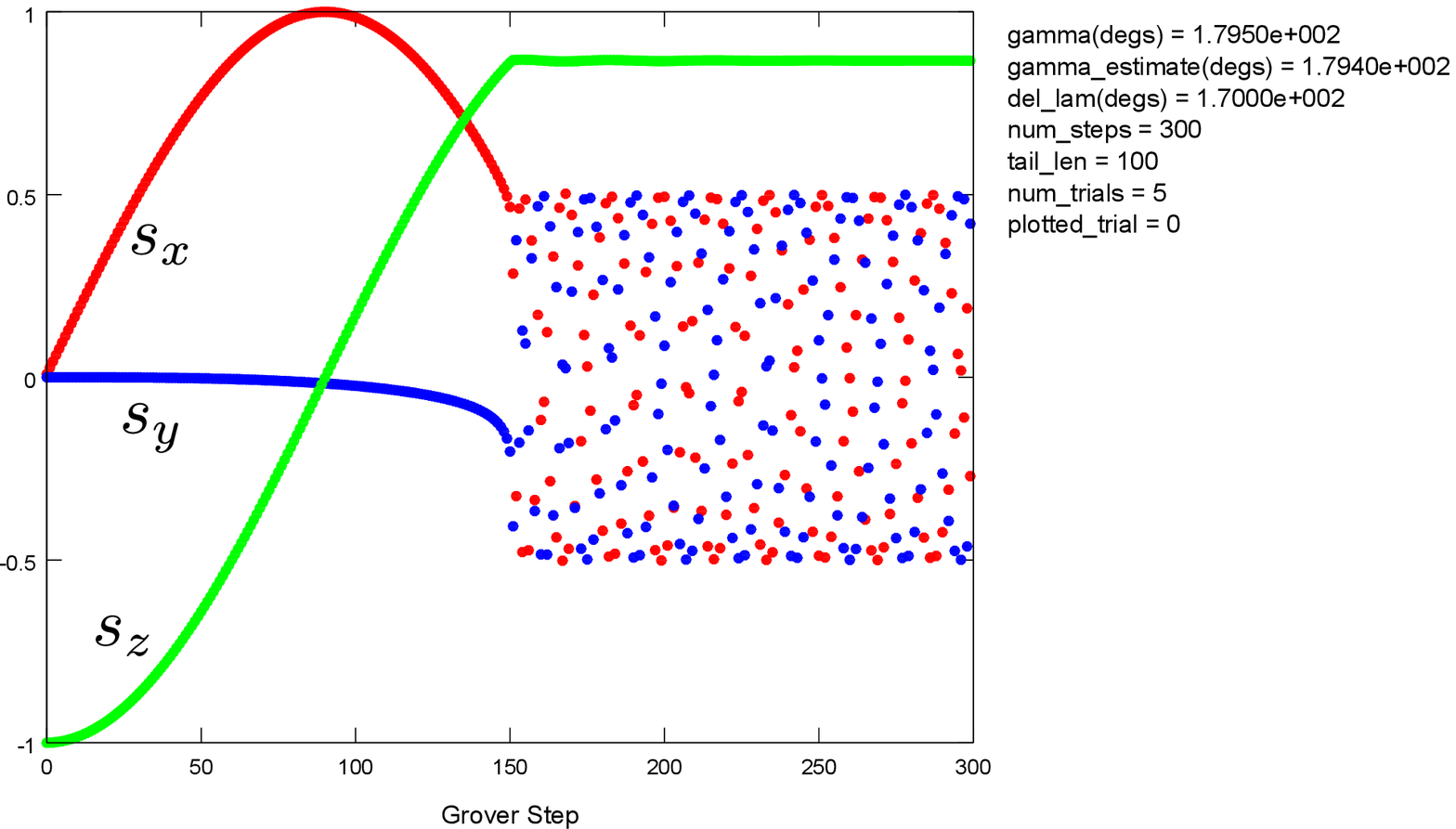, width=5in}
    \caption{
    The 3 components of the unit vector $\hat{s}_j$
    as a function of the Grover step $j$. Plot generated by {\tt afga\_blind.m} with indicated inputs.
    }
    \label{fig-afga-blind179-0}
    \end{center}
\end{figure}

\begin{figure}[b]
    \begin{center}
    \epsfig{file=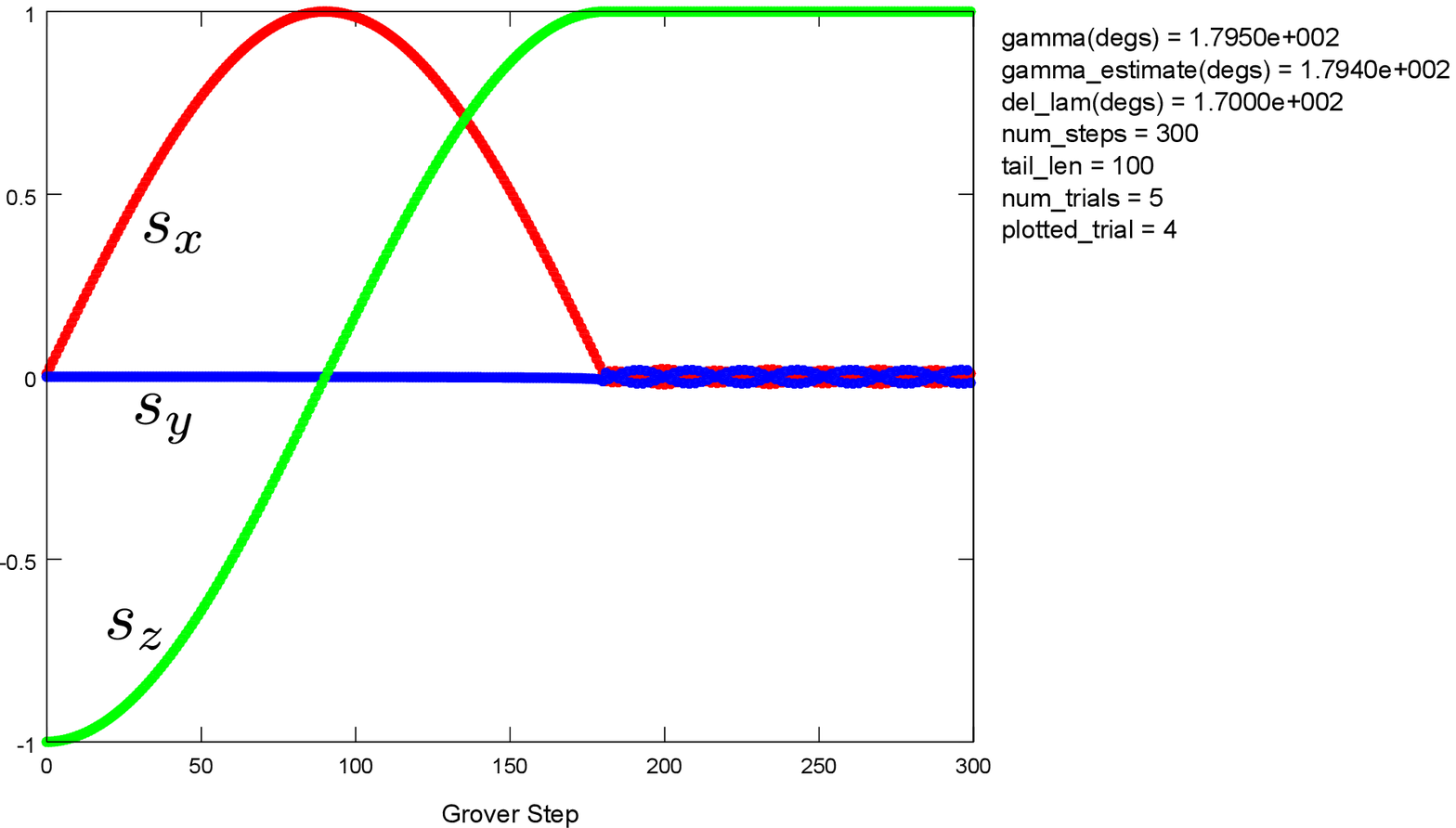, width=5in}
    \caption{
    The 3 components of the unit vector $\hat{s}_j$
    as a function of the Grover step $j$. Plot generated by {\tt afga\_blind.m} with indicated inputs.
    }
    \label{fig-afga-blind179-4}
    \end{center}
\end{figure}

Here are some sample plots.
\begin{itemize}
\item
We got Fig.\ref{fig-afga-blind90-0}
with {\tt plotted\_trial}=0 (first trial)
and with a $\gamma_0$ close to 90 degrees.
Then we changed {\tt plotted\_trial}
to 1 and got Fig.\ref{fig-afga-blind90-1}.
\item
We got Fig.\ref{fig-afga-blind179-0}
with {\tt plotted\_trial}=0 (first trial)
and with a $\gamma_0$ close to 180 degrees.
Then we changed {\tt plotted\_trial}
to 4 and got Fig.\ref{fig-afga-blind179-4}.
\end{itemize}

Further plots can be generated by the user using {\tt afga\_blind.m}.
Note that $\gamma_0=180-\epsilon$ degrees,
where $0<\epsilon<<1$,
corresponds to the regime $|\av{t|s}|<<1$
of the ``hardest" problems.
In that
regime of hardest problems,
we found that
larger $N_{Gro}$ and larger $N_{tail}$ are
required for convergence
than in other regimes.
Furthermore, in this regime
the algorithm becomes very sensitive
to various adjustable input parameters like $N_{Gro}$, $N_{tail}$, $\Delta \lam$  and to
the type of low pass filter we use.
We used two types of low pass filters
in the software. The user can test them
both himself.
One was the MMM filter; i.e., a ``min-max-mean" filter that
uses $[\max_r(\hat{s}_r) + \min_r(\hat{s}_r)]/2$.
We found the MMM filter
to be the more robust of the two filters
we tried.
The example plots presented in this section of the paper
were all generated using the MMM filter.
Further work will be required to
determine how to choose
adjustable input parameters
and a low pass filter which are
optimal, or nearly so, for this
type of algorithm.

\clearpage
\clearpage

\clearpage
\section{Appendix: Method B
of calculating $Q^{(4)}(c^4)$}\label{app-method-b}

In this appendix, we will present
Method B, an alternative to
the Method A that was presented in Sec.\ref{sec-method-a}.
Both methods can be used to calculate $Q^{(4)}(c^4)$.

\begin{figure}[h]
    \begin{center}
    \epsfig{file=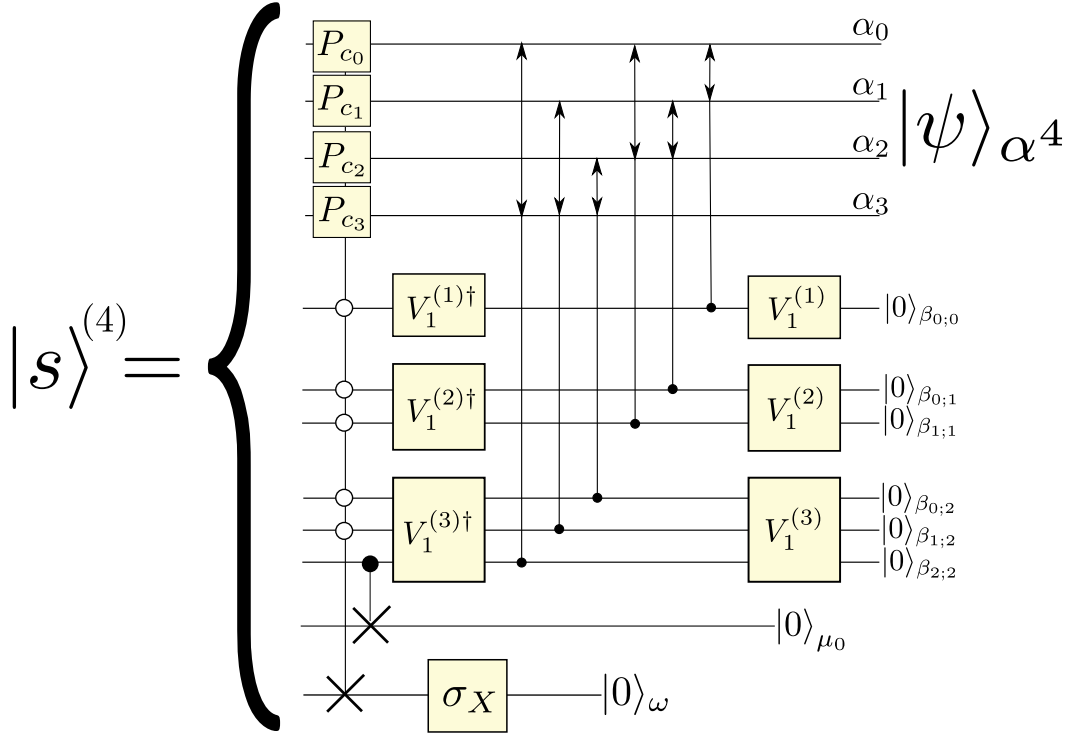, width=5.5in}
    \caption{Method B circuit for
    generating $\ket{s}^{(4)}$
    used in AFGA to calculate
    $|\av{c^4|\pi_{Sym_4}|\psi}|^2$
    }
    \label{fig-sym4-ckt}
    \end{center}
\end{figure}

Unlike in Method A of
calculating
$Q^{(4)}(c^4)$,
in Method B we will
assume the restriction that $\av{c^4|\psi}\geq 0$.
See Appendix \ref{app-restriction}
for cases in which it is
possible to sidestep this restriction.

In Method A,
we applied TTH
(Targeting Two Hypotheses) only once. In method B, we will apply TTH
multiple times, for $k=4,3,2$,
each time applying it
in the way that was described in
Sec.\ref{sec-cats},
together with blind targeting,
and using a sufficient target $\ket{0}_\omega$.
All that remains for
us to do to
fully specify our Method B
circuit for calculating
$Q^{(4)}(c^4)$
is to give a circuit for
generating $\ket{s}^{(k)}$
for $k=4,3,2$.

A circuit for generating
$\ket{s}^{(4)}$ is given by
Fig. \ref{fig-sym4-ckt}.
Note that in this circuit
we do not use the qubit $\gamma$
that was used in method A.
Define $\pi'(\beta)$
to be equal to the
$\pi(\beta)$ defined by Eq.(\ref{eq-def-pi-beta})
but with the projector
$P_0(\beta_{2;2})$ removed.
In other words, ``formally",

\beq
\pi'(\beta)=\pi(\beta)/P_0(\beta_{2;2})
\;.
\eeq
Then Fig.\ref{fig-sym4-ckt}
is equivalent to saying that

\beq
\ket{s}^{(4)}_{\mu,\nu,\omega}=
\sigma_X(\omega)^{
\pi'(\beta)
\pi(\alpha)}
\sigma_X(\mu_0)^{P_1(\beta_{2;2})}
\begin{array}{l}
T(\alpha,\beta)
\begin{array}{r}
\ket{\psi}_{\alpha^4}
\\
\ket{0^6}_\beta
\end{array}
\\
\ket{0}_{\mu_0}
\\
\ket{1}_\omega
\end{array}
\;.
\eeq

\begin{claim}
\label{cl-meth-b-q4}

\beq
\ket{s}^{(4)}_{\mu,\nu,\omega}=
\begin{array}{c}
z_1 \ket{\psi_1}_\mu
\\
\ket{1}_\nu
\\
\ket{0}_\omega
\end{array}
+
\begin{array}{c}
z_0 \ket{\psi_0}_\mu
\\
\ket{0}_\nu
\\
\ket{0}_\omega
\end{array}
+
\begin{array}{c}
\ket{\chi}_{\mu,\nu}
\\
\ket{1}_\omega
\end{array}
\;,
\eeq
for some unnormalized state
$\ket{\chi}_{\mu,\nu}$,
where

\beq
\begin{array}{|c|c|}
\hline
\ket{\psi_1}_\mu=
\begin{array}{l}
\ket{c^4}_\alpha
\\
\ket{0}_{\mu_0}
\end{array}
&
\ket{\psi_0}_\mu=
\begin{array}{l}
\ket{c^4}_\alpha
\\
\ket{1}_{\mu_0}
\end{array}
\\
\ket{1}_\nu=
\left[
\begin{array}{r}
\ket{0}_{\beta_{;0}}
\\
\ket{00}_{\beta_{;1}}
\\
\ket{000}_{\beta_{;2}}
\end{array}
\right]
&
\ket{0}_\nu=
\left[
\begin{array}{r}
\ket{0}_{\beta_{;0}}
\\
\ket{00}_{\beta_{;1}}
\\
-\ket{100}_{\beta_{;2}}
\end{array}
\right]
\\
\hline
\end{array}
\;,
\eeq

\begin{subequations}
\label{eq-z0-z1-new-constraints}
\beq
z_1 = \sqrt{Q^{(4)}(c^4)}\geq 0
\;,
\eeq

\beq
z_0 +z_1= \frac{2}{4}\sqrt{Q^{(3)}(c^4)}
\;,
\eeq
\end{subequations}

\begin{subequations}
\label{eq-z0-z1-old-constraints}
\beq
\frac{|z_0|}{|z_1|} = \sqrt{\frac{P(0)}{P(1)}}
\;,
\eeq

\beq
sign(z_0)=
\left\{
\begin{array}{l}
+1 \mbox{ if } P(+)>P(-)
\\
-1 \mbox{ otherwise }
\end{array}
\right.
\;
\eeq
\end{subequations}
\end{claim}
\proof

\begin{figure}[h]
    \begin{center}
    \epsfig{file=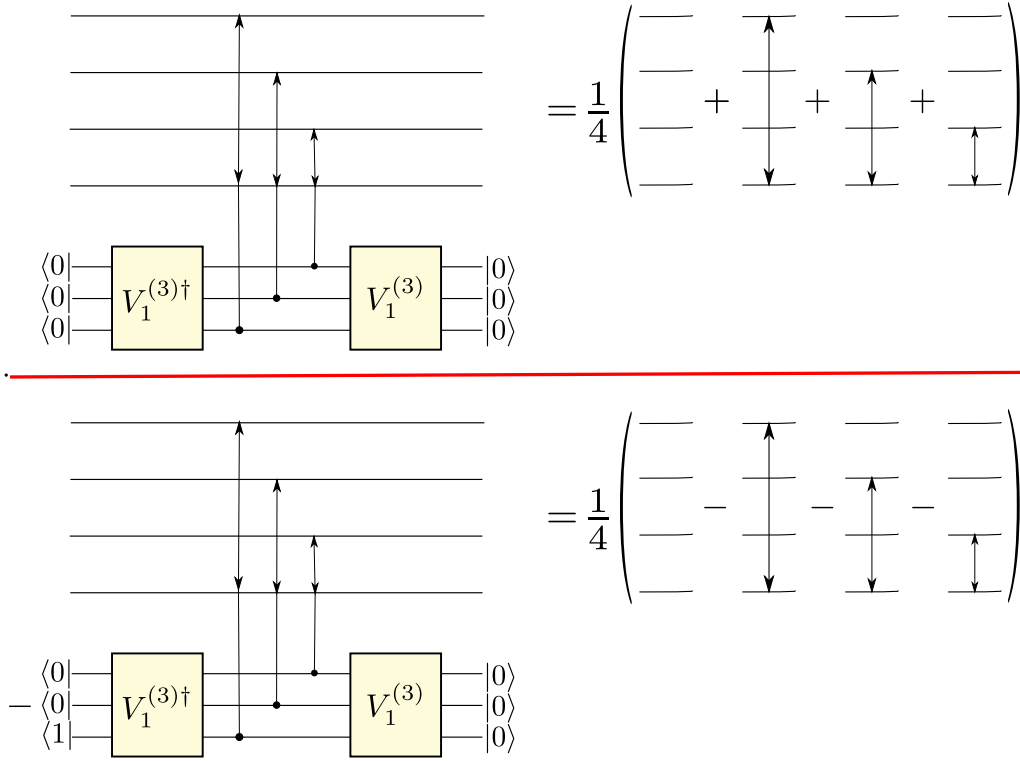, width=5.5in}
    \caption{Two matrix elements
    of $\beta_{;2}^3$.
    }
    \label{fig-v3-ckt}
    \end{center}
\end{figure}

According to Claims \ref{cl-cs-values-4} and
\ref{cl-cs-values-4-neg},

\beq
V_1^{(3)}
\begin{array}{c}
\ket{0}
\\
\ket{0}
\\
\ket{0}
\end{array}
=
\frac{1}{\sqrt{4}}
\left[
\begin{array}{c}
\ket{0}
\\
\ket{0}
\\
\ket{0}
\end{array}
+
\begin{array}{c}
\ket{1}
\\
\ket{0}
\\
\ket{0}
\end{array}
+
\begin{array}{c}
\ket{0}
\\
\ket{1}
\\
\ket{0}
\end{array}
+\begin{array}{c}
\ket{0}
\\
\ket{0}
\\
\ket{1}
\end{array}
\right]
\;,
\eeq
and

\beq
V_1^{(3)}
\begin{array}{c}
\ket{0}
\\
\ket{0}
\\
\ket{1}
\end{array}
=
\frac{1}{\sqrt{4}}
\left[
-
\begin{array}{c}
\ket{0}
\\
\ket{0}
\\
\ket{0}
\end{array}
+
\begin{array}{c}
\ket{1}
\\
\ket{0}
\\
\ket{0}
\end{array}
+
\begin{array}{c}
\ket{0}
\\
\ket{1}
\\
\ket{0}
\end{array}
+\begin{array}{c}
\ket{0}
\\
\ket{0}
\\
\ket{1}
\end{array}
\right]
\;.
\eeq
Therefore,
Figure \ref{fig-v3-ckt}
is true. Figs.\ref{fig-sym4-ckt} and \ref{fig-v3-ckt} imply
the two constraints
given by Eqs.(\ref{eq-z0-z1-new-constraints}).

The two constraints given by
Eqs.(\ref{eq-z0-z1-old-constraints})
are old news. They come directly
from Eq.(\ref{eq-z1-z0})
and Eq.(\ref{eq-z01-cos-theta}).
Note that
 $\cos \theta = sign(z_0)$ for the special case being considered
 in this claim, namely when $z_1$ is non-negative real and $z_0$ is either positive or
negative real.
\qed

\begin{figure}[h]
    \begin{center}
    \epsfig{file=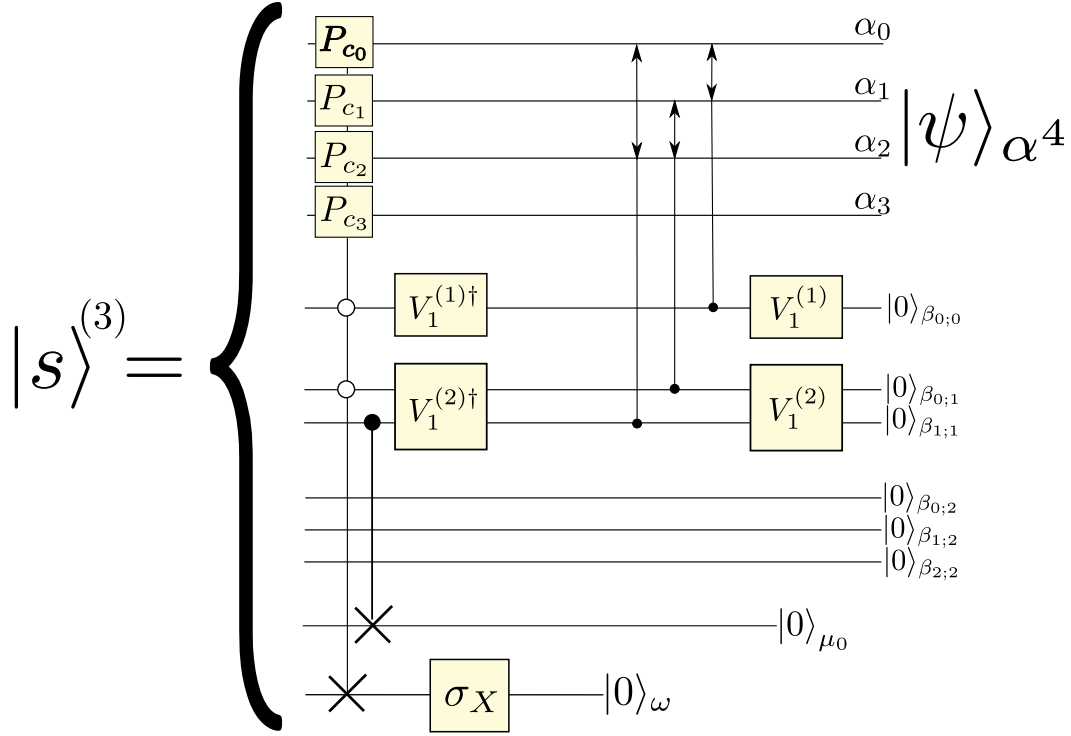, width=5.5in}
    \caption{Method B circuit for
    generating $\ket{s}^{(3)}$
    used in AFGA to calculate
    $|\av{c^4|\pi_{Sym_4}|\psi}|^2$
    }
    \label{fig-sym3-ckt}
    \end{center}
\end{figure}

After doing TTH with
$\ket{s}^{(4)}$,
we are left knowing
$Q^{(4)}(c^4)$ in terms of
$Q^{(3)}(c^4)$. If we know
$Q^{(3)}(c^4)$, we can stop right there
and we are done.
Otherwise, we can
do TTH again, this time
with the $\ket{s}^{(3)}$
given by Fig.\ref{fig-sym3-ckt}.
Again, we are left knowing
$Q^{(3)}(c^4)$ in terms of
$Q^{(2)}(c^4)$. If we know
$Q^{(2)}(c^4)$, we can stop right there
and we are done.
Otherwise, we can
do TTH again, this
time with the $\ket{s}^{(2)}$
given by a circuit
analogous to Figs.\ref{fig-sym4-ckt}
and \ref{fig-sym3-ckt}.
Eventually we end up
finding $Q^{(4)}(c^4)$
in terms $Q^{(1)}(c^4)$. We
assume the latter is known.

\begin{claim}
\beq
\sqrt{Q^{(4)}(c^4)}=
\frac{
\frac{2}{4}
\frac{2}{3}
\frac{2}{2}
\sqrt{Q^{(1)}(c^4)}
}{
\prod_{k=4,3,2}
\left\{
1 + \sigma^{(k)}
\sqrt{\frac{P^{(k)}(0)}{P^{(k)}(1)}}
\right\}
}
\;
\eeq

\beq
\sigma^{(k)}=
\left\{
\begin{array}{l}
+1 \mbox{ if } P^{(k)}(+)>P^{(k)}(-)
\\
-1 \mbox{ otherwise}
\end{array}
\right.
\;
\eeq
\end{claim}
\proof
Follows from Claim \ref{cl-meth-b-q4}, by analogy.
\qed

\section{Appendix: Linear
Transform of Vector If Vector
Not Normalized}
\label{app-restriction}

Often, when calculating with a
quantum computer the linear transform
of a vector $\ket{\psi}$,
our algorithm works only if we assume that the vector $\ket{\psi}$ has non-negative components in some basis,
or is normalized in some norm, or both.
The purpose of this appendix is to
show that this restriction on
$\ket{\psi}$ does
not imply a large reduction of generality
of the algorithm. We will show that
given some simple
information about $\ket{\psi}$,
we can still use the restricted algorithm
to
find the  linear transform of $\ket{\psi}$,
even if $\ket{\psi}$ doesn't satisfy the
restrictions. The results of this appendix
are very obvious but worth keeping in mind.

For any $z\in \CC$, let
$z_r,z_i$ be its real and imaginary parts respectively.

We wish
to consider some finite
 set $S_\rvx$ and
 two functions $f, f^-:S_\rvx\rarrow \CC$ related by

\beq
f(x)=\sum_{x^-\in S_\rvx}
M(x,x^-)f^-(x^-)
\;
\eeq
where $M(x,x^-)\in \CC$.
Function $f$ will be
referred to as the M-transform of
function $f^-$.

\begin{claim}
If one is given constants $a_r,a_i,b_r, b_i\in \RR$
such that

\beq
a_r<f^-_r(x^-)<b_r
\;,\;\;
a_i<f^-_i(x^-)<b_i
\;
\eeq
for all $x^-$,
and one is given $\sum_{x^-} M(x,x^-)$, then
the M-transform of $f^-(x^-)$
can be calculated easily from
the M-transform of functions
$g_r(),g_i()$
which satisfy
$0\leq g_r(x^-),g_i(x^-)\leq 1$ for all $x^-$.
\end{claim}
\proof
Define
\beq
L=\max(b_r-a_r, b_i-a_i)
\;
\eeq
and

\beq
g_r(x^-) = \frac{f^-_r(x^-)-a_r}{L}
\;,\;\;
g_i(x^-) = \frac{f^-_i(x^-)-a_i}{L}
\;.
\eeq
Then

\beq
\sum_{x^-} M(x,x^-)
\left[\frac{f^-(x^-)-a}{L}\right]
=
\sum_{x^-} M(x,x^-)g^-_r(x^-)
+
i\sum_{x^-} M(x,x^-)g^-_i(x^-)
\;.
\eeq
\qed

\begin{claim}
If one is given a constant $N>0$, then
the M-transform of $f^-(x^-)$
can be easily calculated from the M-transform of $f^-(x^-)/N$.
For instance, $N$ might be
$\sqrt{\sum_{x^-}|f^-(x^-)|^2}$.
\end{claim}
\proof
\beq
f(x)=N\sum_{x^-\in S_\rvx}
M(x,x^-)\frac{f^-(x^-)}{N}
\;.
\eeq
\qed

\end{document}